\documentclass[aps,prb,twocolumn,superscriptaddress,showpacs,floatfix]{revtex4-2}
\usepackage{comment}
\usepackage{physics}
\usepackage{amsmath,amsthm,amssymb,amsfonts, color, comment, graphicx}
\usepackage{xcolor}
\usepackage{float}
\usepackage{graphicx}
\usepackage{dcolumn}% Align table columns on decimal point
\usepackage{bm}% bold math
\UseRawInputEncoding 
\usepackage{dsfont}
\usepackage{hyperref}% add hypertext capabilities

\newcommand{\be}{\begin{equation}}
\newcommand{\ee}{\end{equation}}

\newcommand{\bQ}{{{\bf{Q}}}}

\newcommand{\br}{{\bf{r}}}

\newcommand{\bq}{{\bf{q}}}

\newcommand{\beal}{\begin{align}}
\newcommand{\eeal}{\end{align}}
\newcommand{\ra}{\rangle}
\newcommand{\la}{\langle}

\renewcommand{\vec}[1]{\mathbf{#1}}

\def\bS{\vec S}

\usepackage{graphicx}
\usepackage[caption=false]{subfig}

\begin{document}

\preprint{APS/123-QED}
\title{$SU(N)$ spin-phonon simulations of Floquet dynamics in spin $S > 1/2$ Mott insulators}
%\title{Spin-phonon Floquet dynamics}
% Force line breaks with \\
\author{Ruairidh Sutcliffe}
\thanks{These authors contributed equally to this work.}
\affiliation{Department of Physics, University of Toronto, 60 St. George Street, Toronto, ON, M5S 1A7 Canada}
\author{Kathleen Hart}
\thanks{These authors contributed equally to this work.}
\affiliation{Department of Physics, University of Toronto, 60 St. George Street, Toronto, ON, M5S 1A7 Canada}
\author{Gil Refael}
\affiliation{Department of Physics, California Institute of Technology, Pasadena CA 91125, USA}
\affiliation{Institute for Quantum Information and Matter, California Institute of Technology, Pasadena CA 91125, USA}
\author{Arun Paramekanti}
%\email{arun.paramekanti@utoronto.ca}
\affiliation{Department of Physics, University of Toronto, 60 St. George Street, Toronto, ON, M5S 1A7 Canada}

\date{\today}

\begin{abstract}
The dynamics of magnetic moments coupled to phonons is of great interest for understanding spin transport
in solids as well as for our ability to control magnetism via tailored phonon modes. For spin $S > 1/2$, spin-orbit coupling permits
an unusual {\em{linear}} coupling of phonons to quadrupolar moments, so that phonons act as a dynamical transverse
field for the spins. Here, we develop a generalized $SU(N)$ spin-phonon Monte Carlo and molecular dynamics technique
to simulate the equilibrium and nonequilibrium properties of such spin-orbital-phonon coupled Mott insulators, and
apply it to a spin-1 model with competing XY antiferromagnet 
(AFM) and quadrupolar paramagnet (QPM) phases. We uncover a rich 
set of nonequilibrium phenomena from driven phonons, including the generation of 
a uniform magnetization in the QPM and AFM, strengthening of N\'eel order, gapping of the AFM Nambu-Goldstone mode by 
Floquet-Ising anisotropy,
and creating Floquet copies of transverse and longitudinal spin waves. {Our work
is relevant for driven spin-1 magnets, such as $\rm{Ba_2FeSi_2O_7}$, and we highlight broader implications for
nonequilibrium multipolar magnetism.}
% We then examine the effect of dynamical excitation of these phonon modes in both pump-probe and continuous driving experiments in a model hosting a  pseudospin-$1/2$ non-Kramers doublet and a spin-1 model. In the pseudospin-$1/2$ model we look at the effect of phenomenological damping in the phonon sector in addition to providing evidence for octupolar order switching occurring at domain walls. We find that the damping strength can prevent octupolar order switching by restricting the amplitude of the phonon modes. In the spin-1 model, we show, in equilibrium, the onset of an ${\cal E}_g$ distorted state by tuning the spin-phonon coupling strength.
% Further, we model the effect of continuous, resonant, excitation of a single ${\cal E}_g$ phonon mode and show how, through renormalization of the model parameters, such a drive is able to induce antiferromagnetic order, and possibly, an out-of-equilibrium phase transition.
\end{abstract}
`\maketitle

\section{Introduction}
\label{sec:Intro}

% {\it Introduction.---} 
The interaction between lattice and electronic degrees of freedom in quantum materials gives rise to a 
rich vein of interesting equilibrium phenomena in solids including the Jahn-Teller effect \cite{Khomskii_2014}, polaron formation
\cite{cesare2021polarons,alexandrov2010advances,zhang2023_bipolaron,Han2024_PRLbipolaron}, and 
superconductivity in conventional materials as well as possibly in Moir\'e materials
\cite{Bardeen1973_superconductivity,marsiglio2008electron,carbotte1990properties,Wu2018_PRLtwistedphonon}.
Recent years have witnessed the exploration of non-equilibrium effects in such systems, boosted by advances in 
optical and THz sources which have enabled ultrafast resonant excitation of optical phonon modes
\cite{stupakiewicz2021ultrafast,schnyder2011resonant,forst2011nonlinear,mashkovich2021terahertz,Intro_Cavalleri_nonlinearTHZ}. 
These developments have proved important for pump-probe spectroscopy of solids as well as for the creation of remarkable
Floquet-driven states of electronic matter \cite{Intro_Floquet_Fiete_AnnPhys2021,Intro_Floquet_Oka_ARCMP2019,Intro_floquet_Lindner_PRB2013}.
The ability to resonantly drive lattice vibrations has enabled the exploration of non-equilibrium
% Experiments and theory have explored 
signatures of superconductivity or strongly enhanced electron mobility
\cite{Intro_Cavalleri_HiTcPhonon_PRX2020,dodge2023_PRLspurious}, the stabilization of ferroelectric order
\cite{Intro_Cavalleri_ferroelec_PRL2017}, generation of
large effective magnetic fields from chiral-phonon drive in dipolar magnets \cite{Intro_ChiralPhonon_chaudhary2023giant, juraschek2022giant,hernandez2022chiral,nova2017effective,afanasiev2021ultrafast,stupakiewicz2021ultrafast,davies2024phononic}, and control of magnetism and dynamic Kerr rotation in
two-dimensional (2D) van der Waals magnets \cite{Afanasiev_Science2021,Padmanabhan_NatComms2022}.

More recent work has begun to explore the physics of phonons in \emph{multipolar} materials where 
spin-charge-orbital entangled degrees of freedom such as quadrupoles, octupoles, or higher order multipoles 
can lead to distinct forms of broken symmetry phases. In particular, time-reversal-even multipoles in such
materials can {\em linearly} couple to symmetry-compatible lattice modes, so that judiciously 
applied strain fields \cite{Maharaj2017_strain,Ikeda2018_strain,patri2019unveiling,patri2020theory,patri2020distinguishing,ye2023_octupolestrain,voleti2023probing} 
or dynamical phonon excitations \cite{hart2024phonon} can be used to probe hidden multipolar orders and their 
dynamics. 
Further, recent THz experiments have found evidence that phonon-multipolar coupled dynamics can lead to nonequilibrium effects, 
like triplon spin excitations in a dimer magnet \cite{Intro_ShastrySutherland_Ruegg_PRB2023}, and even phonon-driven fluctuations into a hidden quadrupolar-ordered state in $\rm{Ca_2RuO_4}$ \cite{ning2023coherent}.
% Recent THz experiments have found evidence for phonon-multipolar coupled dynamics \cite{Intro_ShastrySutherland_Ruegg_PRB2023}, 
% and revealed phonon-driven fluctuations into a hidden quadrupolar-ordered state in $\rm{Ca_2RuO_4}$ \cite{ning2023coherent}.

\begin{figure}[t]
\centering
\includegraphics[width=0.49\textwidth]{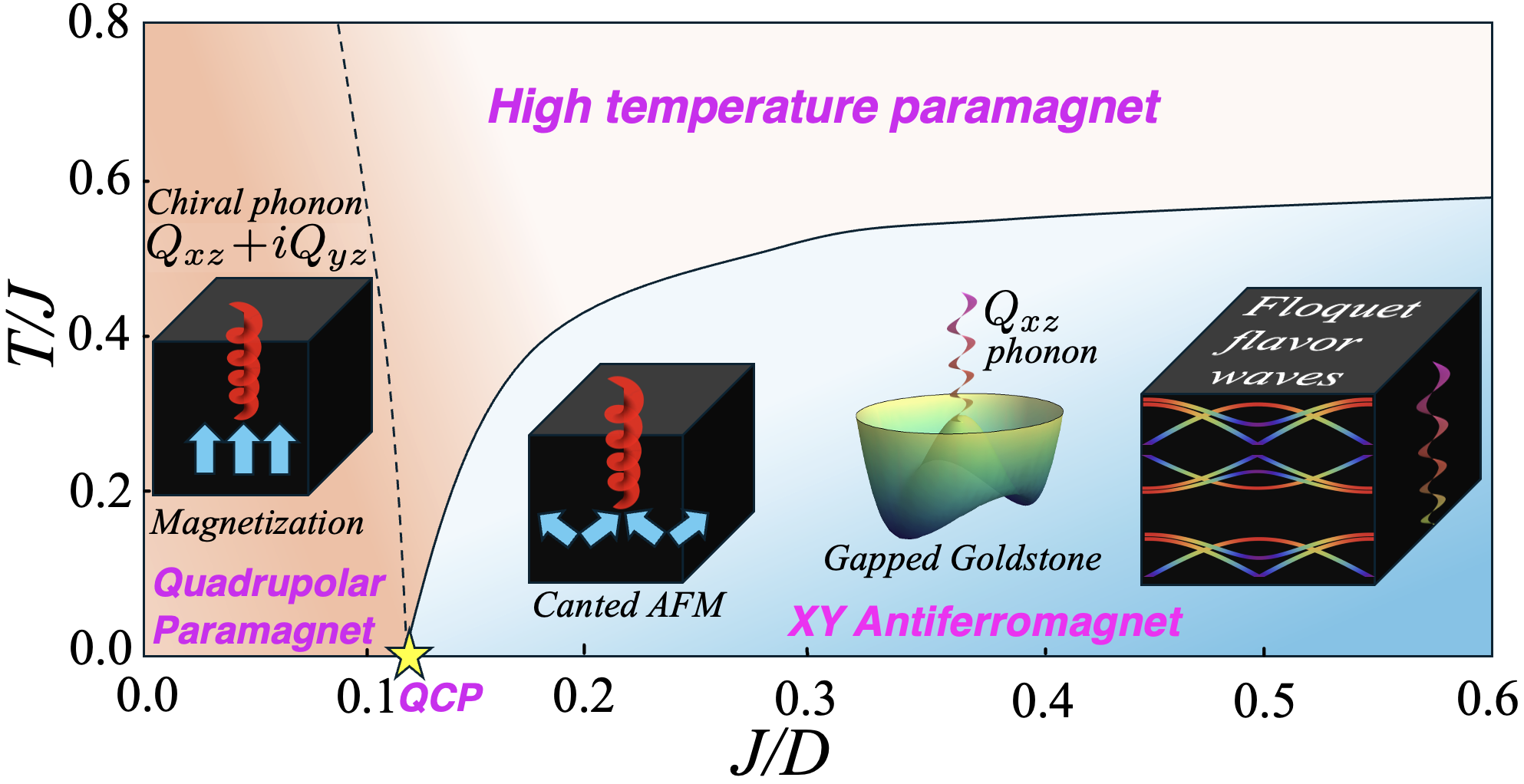}
\caption{Monte Carlo phase diagram of the spin-$1$ model on the tetragonal lattice with single-ion anisotropy $D$ and Heisenberg 
exchange $J$ including moderate spin-phonon coupling to $(Q_{xz},Q_{yz})$ phonon modes as described in the text. At $T\!=\!0$, there 
is a quantum critical point (QCP) separating the
quadrupolar paramagnet (QPM) with $S_z\!=\! 0$ from an XY N\'eel antiferromagnet (AFM). $\rm{Ba_2FeSi_2O_7}$ sits in the
AFM phase at $J/D \!\approx\!0.17$. Solid and dashed line respectively
indicate computed phase transition and crossover into the high temperature paramagnet. 
Our focus here is on the highlighted
non-equilibrium Floquet phenomena achieved by driven-dissipative phonon modes in this multipolar magnet.}
% including nonzero magnetization on the QPM, a spin canted AFM, 
% gapped Goldstone modes, and higher Floquet copies of dipolar and quadrupolar flavor waves.}
% find a quantum phase transition at $J/D\!\approx\!0.18$ separating a quadrupolar paramagnet (QPM) with $S_z\!=\!0$ from
% a N\'eel ordered antiferromagnet (AFM), with a classical paramagnet (CPM) from thermal disordering of the AFM.
% Solid and dashed lines indicate phase transition and crossover respectively. Color scale shows the average spin-dipole length in the
% different phases}
\label{fig:PhaseDiagram}
\end{figure}

Multipolar magnetism emerges naturally in systems with effective pseudospin $S \!> \! 1/2$;
for instance, the set of local observables for a pseudospin $S\!=\!1$ magnet includes five quadrupole operators 
in addition to the three $SU(2)$ generators which are dipole operators. The
equilibrium and temporal correlations of all multipole operators for a spin-$S$ magnet 
can be captured by an $SU(N)$ generalization of the conventional Monte Carlo method and Landau-Lifshitz
dynamics \cite{stoudenmire2009quadrupolar,dahlbom2022langevin,remund2022semi}, which recasts all multipole 
operators for a pseudospin-$S$ system as linear combinations of the $N^2\!-\!1$
generators of the $SU(N)$ algebra where $N\!=\!2S\!+\!1$.
% with the expectation value of any 
% multipole being computed using the corresponding spin wavefunction which is an $N$-dimensional 
% complex vector. 
This $SU(N)$ approach has been used to successfully describe effective $S\!=\! 1$ magnets
NiGa$_2$S$_4$ \cite{stoudenmire2009quadrupolar} and $\rm{Ba_2FeSi_2O_7}$ \cite{do2021decay,do2023understanding}.
{Given the diversity of materials exhibiting spin-phonon coupling,
it is important to generalize this method to incorporate phonon degrees of freedom
and to extend it to the hitherto unexplored regime of non-equilibrium dynamics.}

% Conventional Monte Carlo (MC) methods and semiclassical Landau-Lifshitz equations, which are grounded in the 
% $SU(2)$ coherent state representation of spins, are unable to capture these quadrupolar observables.
% %Indeed, \emph{no} $SU(2)$ rotation can take a spin coherent state, such as $|1\ra$, to a 
% %multipolar eigenstate such as $(|1\ra + |-1\ra)/\sqrt{2}$. 
% This hurdle has been overcome by recognizing that
% all multipole operators for a pseudospin-$S$ system can be expressed as linear combinations of the $N^2\!-\!1$
% generators of the $SU(N)$ algebra where $N\!=\!2S\!+\!1$, and the expectation values of any multipole
% can be computed using the spin wavefunction which is an $N$-dimensional complex vector. MC sampling of these 
% wavefunction amplitudes or computing their Schr\"odinger time evolution allows one to capture the 
% thermodynamics and dynamic structure factor of the
% all multipolar degrees of freedom on equal footing 

{This paper reports substantial progress on both issues: we extend the $SU(N)$ approach to 
incorporate weakly dissipative phonons and pseudospin-phonon coupling, and we apply this method to
explore the Floquet dynamics of spin-$S$ magnets. We illustrate the power of this technique by highlighting the 
rich physics of driven spin-phonon coupled dynamics in a three-dimensional (3D) $S\!=\!1$ Heisenberg model with single-ion 
anisotropy.}
%which describes the low energy magnetism in $\rm{Ba_2FeSi_2O_7}$.
In equilibrium, and 
for weak to moderate spin-phonon coupling, this model exhibits two phases, a quadrupolar paramagnet (QPM) 
and an ordered XY antiferromagnet (AFM), separated by a critical point describing
Bose-Einstein condensation (BEC) of $S_z=\pm 1$ excitons.
Our study reveals remarkable out-of-equilibrium phenomena in this system induced by resonant one-phonon and two-phonon 
drives in the QPM and ordered AFM as highlighted in Fig.~\ref{fig:PhaseDiagram}. 

Studying a coherent chiral phonon drive in the QPM and AFM, we discover steady states 
with weak induced ferromagnetic order and a QPM-AFM transition.
%. Our approach goes beyond previous work on chiral-phonon induced effective magnetic fields 
% in $4f, 3d$ magnets in providing access to real-time dynamics and spatiotemporal correlations
% on the lattice. 
We also show that a one-phonon drive
in the AFM leads to uniaxial strengthening of XY AFM order resulting in a gapped
Nambu-Goldstone mode in the dynamical spin structure factor, and to higher Floquet
copies of the spin modes. We shed light on our steady state observations
using a Floquet-Magnus expansion. {While we focus here on a concrete model 
potentially relevant to $\rm{Ba_2FeSi_2O_7}$, where our predictions could be tested in future experiments, 
our work is easily generalizable to any spin-$S$ model. This sets the stage for exploring 
non-equilibrium dynamics in a wide class of multipolar magnets.}

\section{SU(N) spin-phonon simulations}
\label{sec:SUN}
% \noindent {\it SU(N) spin-phonon simulations.---}
%Vibronic coupling in conventional magnets has long been a rich source of interesting and novel physics. 
%On symmetry grounds, phonons cannot directly couple to the time-reversal-odd spin dipole moment. 
In conventional dipolar
magnets, Einstein phonons  modulate the bilinear spin-exchange interaction, potentially stabilizing
unusual orders in geometrically frustrated systems \cite{bergman_prb2006,wang2008spin,szasz2022phase,watanabe2023magnetic}. 
By contrast, Einstein phonons in multipolar magnets can directly couple to on-site quadrupolar moments of matching symmetry.
%We develop simulation techniques to study the equilibrium and non-equilibrium properties of such multipolar magnets.
% We write the local Einstein phonon Hamiltonian
% \begin{equation}
% H_{\rm{ph}}=\sum_{\br\alpha} \left( \frac{1}{2}M_\alpha \Omega_\alpha^2Q_\alpha^2(\br) +\frac{P_\alpha^2(\br)}{2M_\alpha}\right),
% \label{eq:ph}
% \end{equation}
% where $\alpha$ indexes the phonon,  $\Omega_\alpha$ is the phonon frequency and $M_\alpha$ is the phonon mass, and express
% the symmetry allowed spin-phonon coupling Hamiltonian as
% \begin{equation}
% H_{\rm{sp-ph}}=- \lambda \sum_{\br,\alpha}  Q_{\alpha} (\br) T_\alpha (\br)
% \label{eq:sp-ph}
% \end{equation}
% Here the phonons and multipoles are chosen to transform under a single irreducible point group representation,
% %Here, $\Gamma$ indexes the representation{\bf unclear - should it also have a gamma on the T? Also, why not enumerate the first two equations?}, 
% $\alpha$ indexes the basis functions of this representation, and $T_\alpha$ are quadrupolar moment operators (expressible
% using spin operators);
% this approach is easily generalizable to multiple point group representations.
% The quadrupole-phonon coupling strengths $\lambda$ may be obtained via
% density functional theory.
% The full Hamiltonian of interest is then $H = H_{\rm sp}+H_{\rm{ph}} + H_{\rm{sp-ph}}$
% where $H_{\rm sp}$ encodes intersite pseudospin dipolar and multipolar exchange couplings.
The Einstein phonon and symmetry allowed spin-phonon Hamiltonians
are given by
\begin{eqnarray}
H_{\rm{ph}}&=&\sum_{\br,\alpha} \left( \frac{1}{2}M_\alpha \Omega_\alpha^2Q_\alpha^2(\br) +\frac{P_\alpha^2(\br)}{2M_\alpha}\right),
\label{eq:ph} \\
H_{\rm{sp-ph}}&=&- \lambda \sum_{\br,\alpha}  Q_{\alpha} (\br) T_\alpha (\br)
\label{eq:sp-ph}
\end{eqnarray}
Here the phonons and multipoles are chosen to transform under a single irreducible point group representation, and
%Here, $\Gamma$ indexes the representation{\bf unclear - should it also have a gamma on the T? Also, why not enumerate the first two equations?}, 
$\alpha$ indexes the basis functions of this representation. ($Q_\alpha$, $P_\alpha$) denote the phonon coordinate and momentum, respectively, $\Omega_\alpha$ is the phonon frequency, $M_\alpha$ is the phonon mass, and $T_\alpha$ are quadrupolar moment operators (expressible
using spin operators).
%this approach is easily generalizable to multiple point group representations.
The quadrupole-phonon coupling strengths $\lambda$ may be obtained via
density functional theory.
The full Hamiltonian of interest is then $H = H_{\rm sp}+H_{\rm{ph}} + H_{\rm{sp-ph}}$
where $H_{\rm sp}$ encodes intersite pseudospin dipolar and multipolar exchange couplings.

% The simulations in this work are generally termed $SU(N)$, where $N=2S+1$, for a spin-S system, since one can track the expectation values of all $N^2-1$ generators of the representation. From linear combinations thereof, all dipolar and \emph{multipolar} expectation values may be constructed. For $S=1/2$, we have $N=2$ and the three generators of the representation are just the Pauli matrices, which is equivalent to traditional vector Monte Carlo simulations. Using $SU(N)$ spins is advantageous as it allows us to simulate all dipolar and higher order multipolar moments on equal footing.
% For MC updates in these simulations, we update both the spin wavefunctions and phonon coordinates using a Metropolis update scheme.

Our $SU(N)$ spin-phonon simulations extend the work in Ref.\cite{stoudenmire2009quadrupolar,remund2022semi};
%where, in contrast to traditional vector Monte Carlo, 
% where spin expectation values of observables are updated directly, 
at each site we use an $N$-dimensional complex vector for
the local spin wavefunction \emph{and} phase space coordinates $(Q_\alpha(\br),P_{\alpha}(\br))$ for
phonon modes at each site $\br$.
%where $\alpha$ runs over all phonon eigenmodes.
% . Consequently,  Monte Carlo updates are performed, in the spin-$1$ case, by representing the wavefunction of the full lattice as a direct product
%For the $S\!=\!1$ case, corresponding to $N\!=\!3$, 
The spin wavefunction
$|\Psi\rangle=\otimes_\br |\psi_\br\rangle$, with
\begin{eqnarray}
\label{eq:wavefunction}
|\psi_\br \rangle &=& \sum_{m=-S}^S a_m(\br) |m\rangle
\end{eqnarray}
 where $a_m(\br)$ are $N\!=\!2S\!+\! 1$ complex parameters. Wavefunction normalization 
$\sum_m\! |a_m(\br)|^2\!=\!1$ and irrelevance of an overall phase yields $N\!-\!1$ independent complex parameters at each site.
Assuming classical phonons, we integrate
out phonon momenta for equilibrium properties, and
 % and sample only the phonon coordinate for equilibrium properties{\bf the phrase repeats}. 
sample the
set $(\{a_m(\br)\},\{Q_\alpha(\br)\})$ using Monte Carlo techniques to achieve steady state probabilities
 \begin{equation}
     {\cal P} \propto \exp \left[- (\la H_{\rm sp}\ra + \la H_{\rm sp-ph}\ra + H_{\rm ph})/T \right]
 \end{equation}
where $\la.\ra$ denotes the expectation value of Hamiltonian spin operators in the sampled wavefunction.  
To simulate dynamics at any temperature $T$, we start from an
equilibriated MC configuration 
$(\{a_m(\br)\},\{Q_\alpha(\br)\})$,
pick initial phonon momenta from a Boltzmann distribution, and time-evolve this configuration 
to compute spatiotemporal correlations, averaging observables over 
$\sim 200$ initial configurations; see appendix \ref{app:MCdetails} and \ref{app:MDdetails} for simulation
details and \ref{app:tests} for various test cases.

\section{Model and equilibrium phase diagram}
\label{sec:model}
% \noindent {\it Model \& equilibrium phase diagram.---}
As a paradigmatic example of a multipolar magnet we consider a $S\!=\!1$ spin Hamiltonian with dipolar and
quadrupolar degrees of freedom
\begin{equation}
H_{\rm sp} \!=\!\!\sum_{\la \br,\br' \ra}\! 
J_{\br\br'} {\bf S}(\br) \cdot {\bf S}(\br') + D  \sum_{\br} S_{z}^2(\br)
\label{eq:sp}
\end{equation}
where $D$ encodes single-ion anisotropy and $J_{\br\br'}$ is the Heisenberg spin exchange.

For large $D > 0$, this model favors a quadrupolar paramagnet (QPM)
with $S_z\!=\!0$ at each site, as proposed for spin-$1$ Ni$^{2+}$ on the diamond lattice
in NiRh$_2$O$_4$ \cite{diamond_spin1_mcqueen_PRMat2018,diamond_spin1_gangchen_PRB2017}. 
Increasing $J/D$ leads to a magnetically 
ordered phase from BEC of the $S_z=\pm 1$ excitons. 
For instance, the low energy magnetism in $\rm{Ba_2FeSi_2O_7}$, with XY N\'eel order 
coexisting with a quadrupolar moment, is captured by this model on the tetragonal lattice, 
with intralayer $J\!\sim\! 0.26$\,meV, interlayer 
$J' \! \approx \! 0.026\rm{meV}$, and $D \!\sim\! 1.4\rm{meV}$ \cite{do2023understanding}.
{Motivated by this evidence of multipolar magnetism in $\rm{Ba_2FeSi_2O_7}$, we explore
this tetragonal lattice model, fixing $J'/J=0.1$ and varying $J/D$.}

We couple quadrupole moments $T_{\mu\nu}\!=\!(S_\mu S_\nu\!+\!S_\nu S_\mu)$
to phonons $(Q_{xz},Q_{yz})$ (i.e., ${\cal E}_g$ representation for tetragonal symmetry), so that 
$\alpha \! \equiv\! \mu\nu\!=\! xz,yz$ in
Eq.~\ref{eq:ph} and Eq.\ref{eq:sp-ph}.
%via
% \begin{eqnarray}
% H_{\rm{ph}}&=& \!\!\! \sum_{\br,\alpha=xz,yz} \left[ \frac{1}{2} M \Omega^2 Q_{\alpha}^2(\br) + \frac{1}{2 M} P_{\alpha}^2(\br) \right],\\
% H_{\rm sp-ph} &=& \lambda  \sum_\br \left[ Q_{xz}(\br) T_{xz}(\br) + 
% Q_{yz}(\br) T_{yz}(\br) \right],
% \label{eq:spin1ph}
% \end{eqnarray}
%where $T_{\mu\nu}\!=\!(S_\mu S_\nu\!+\!S_\nu S_\mu)$ are quadrupole moments. 
We choose
illustrative parameters $\hbar\Omega/J \!=\! 40$, so that $\hbar\Omega/J \!\gg\! 1$, i.e., phonon oscillations
much faster than spin precession, and fix $\hbar^2/(2 M J a^2) \!=\! 0.03$ where $a$ is the lattice constant.
Henceforth, we set $\hbar\!=\! 1$.
The quadrupole-phonon coupling is set to $\lambda a/J = 350$, so a $1\%$ lattice distortion
results in an energy change $\sim \! J$.
{The precise choice of these parameters
could be extracted from {\it ab initio} calculations; however, we note that our key qualitative 
results discussed below remain robust.}

%We set $\Omega/J_1 = 38$, $\lambda a/J = 2.5$, $\hbar^2/(2 M J_0 a^2) = 80$, $M \Omega^2 a^2/(2J_0) \simeq 4.7$.

{Using $SU(3)$ spin-phonon MC simulations developed in this work, we have determined the equilibrium phase diagram at moderate
spin-phonon coupling} as shown in Fig. \ref{fig:PhaseDiagram}, finding a QPM phase at small $J/D$ and an XY AFM at large $J/D$. This
result is similar to previous work which did not incorporate phonons \cite{do2023understanding}.
$\rm{Ba_2FeSi_2O_7}$ lies in the XY AFM phase at $J/D  \! \approx \! 0.17$, close to the quantum critical point which 
lies at $J/D \! \approx \! 0.12$.
% (see SM \cite{suppmat}). 
{Although phonons do not significantly
impact the equilibrium phase diagram, they provide a knob to control remarkable non-equilibrium dynamics which we
explore below.}

%{\bf the following sentences should be in the next paragraph - no?} \added{done}
% \begin{figure}[t]
% \centering
% \includegraphics[width=0.48\textwidth]{Figures/spin1PhaseDiagram.png}
% \caption{ Equilibrium phase diagram of the spin-$1$ model described by Eq.~(\ref{eq:spin1ham}) and Eq.(\ref{eq:spin1ph}) (see text for details). We
% find a quantum phase transition at $J/D\!\approx\!0.18$ separating a quadrupolar paramagnet (QPM) with $S_z\!=\!0$ from
% a N\'eel ordered antiferromagnet (AFM), with a classical paramagnet (CPM) from thermal disordering of the AFM.
% Solid and dashed lines indicate phase transition and crossover respectively. Color scale shows the average spin-dipole length in the
% different phases.{\bf what is the color scale? What is the distinction between quantum and classical? What is the PD for Barium-Iron?}}
% \label{fig:s1PhaseDiagram}
% \end{figure}

\section{Chiral phonon drive}
% {\it Chiral phonon drive.---}
We begin by studying the impact of a driven chiral phonon generated by pumping two tetragonal 
${\cal E}_g$ phonons with a relative $\pi/2$ phase difference.
Our key result is that the chiral phonon induces a \emph{nonzero} uniform $S_z$ magnetization -
% which can be 
% traced back to a phonon-induced effective magnetic field.
in the AFM, this cants the XY
magnetic order, while in the QPM it induces a non-equilibrium QPM to canted AFM transition.
% Motivated by recent work which shows a phonon-induced quadrupolar state in $\rm Ca_2RuO_4$ \cite{ning2023coherent}, we look at the effect of a coherent drive of both tetragonal ${\cal E}_{g}$ phonons, with a phase difference of $\pi/2$, in the quantum paramagnetic phase. 
\begin{figure}[t]
\label {fig:twoPhononDrive}
\centering
\includegraphics[width=0.48\textwidth]{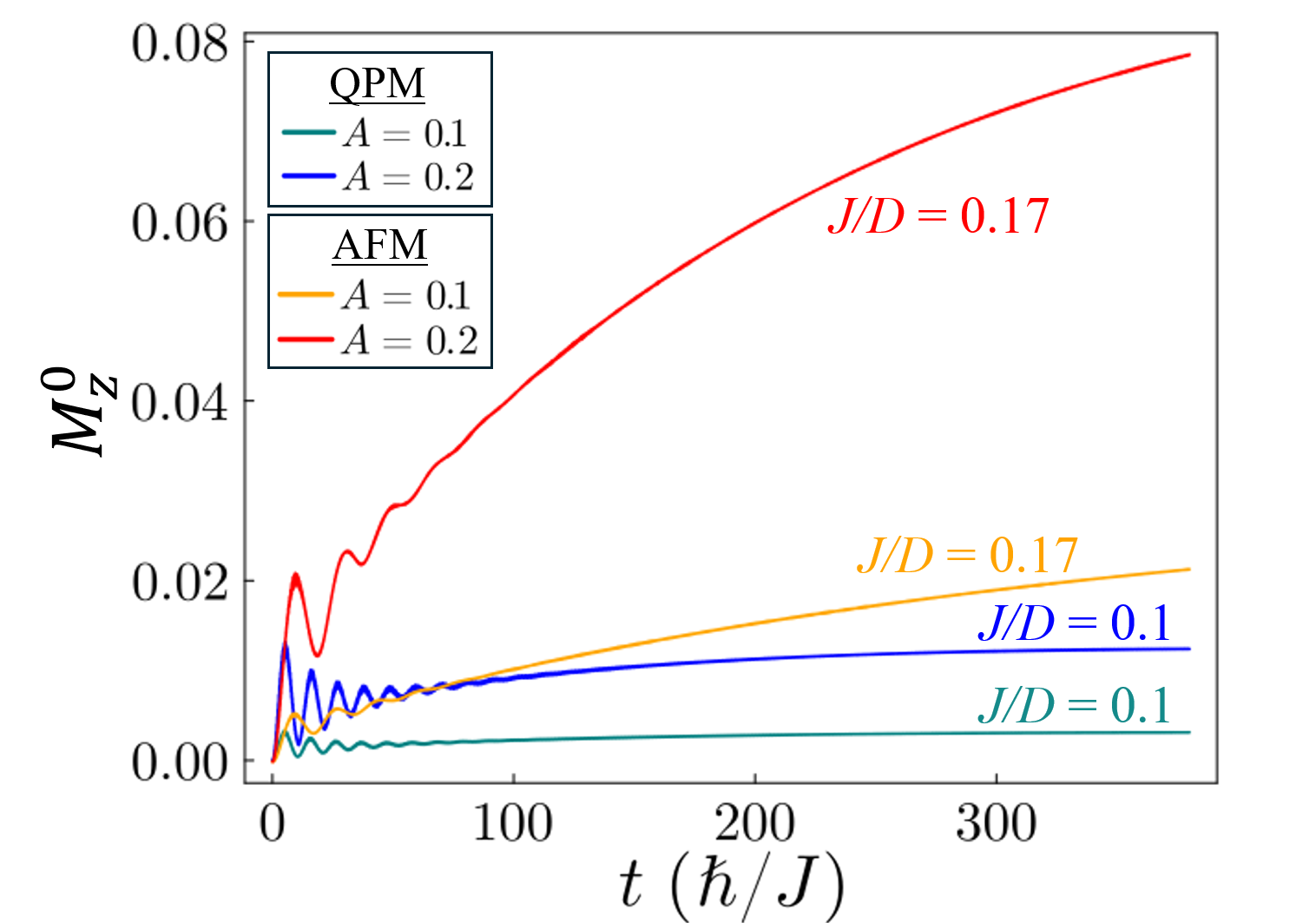}
\caption{ Time evolution of uniform
$S_z$ magnetization, $M^0_z$, induced by a resonant chiral-phonon drive for two different phonon drive strengths $A\!=\!0.1,0.2$, 
at fixed temperature $T/J \!=\! 0.2$ (i) in 
the QPM phase at $J/D\!=\! 0.1$ and (ii) in the AFM phase at $J/D\!=\! 0.17$. 
In both cases, there is a nonzero steady state $M^0_z$ at 
long times, which is enhanced at larger drive strength $A$,
consistent with an effective magnetic field as seen within a lab frame Floquet-Magnus expansion.}
% {\bf add the J/D in the figure itself. Sizes and fonts should match. Could combine the two figs since they look pretty similar.}}
%Heat-map showing the average spin dipole length, $\sqrt{\langle S_x\rangle^2+\langle S_y\rangle^2+\langle S_z\rangle^2}$, for the lattice, as a function of $J/D$ ratio and temperature, T. Overlaid, in white, is the boundary of the non-zero $Q=(\pi,\pi,\pi)$, N\'eel anti-ferromagnetic order. Ultimately, we see at $J/D<\sim 0.16$ a small spin dipole length characteristic of the quantum paramagnetic, QPM, state. Further, as the $J/D$ ratio increases, the system enters the antiferromagnetic (AFM) state, where, initially at small $J/D\sim0.2$, the spin length is shared between the dipole and quadrupole moments before at larger $J/D>0.4$ realizing a nearly pure dipole state that is well approximated by $SU(2)$ coherent states. As temperature increases in all regimes, the system transitions to a classical paramagnet (CPM) wherein the spin length is predominantly in the dipolar sector with a small quadrupolar component.}
\label{fig:twoPhononDrive}
\end{figure}
To study this physics, we incorporate a drive Hamiltonian
$H_{\rm drive}^{\rm ph} = - A \sum_{\br} (Q_{xz}(\br) \cos \Omega t + Q_{yz}(\br) \sin \Omega t)$, where $A$ is the drive amplitude, 
$\Omega$ is resonant with the natural phonon frequency,
and include weak damping such 
that the driven system reaches a steady state. We simulate the full driven-dissipative 
system on the lattice via our $SU(3)$ spin-phonon dynamics technique, tracking the time-dependence of both the uniform
and staggered magnetizations.
Fig.~\ref{fig:twoPhononDrive} shows the chiral-phonon induced uniform $S_z$ magnetization which approaches towards a 
steady state after initial oscillatory transients for various drive strengths in both the QPM and AFM phases.
We observe that the magnetization is induced more easily in the AFM as opposed to the QPM phase, consistent with
a pre-existing significant local dipole moment in the equilibrium AFM.
%\textcolor{red}{?????} In addition, we find a rapidly rotating $S_x(t), S_y(t)$ at
%the phonon frequency which can be captured within a rotating wave approximation \textcolor{red}{?????}.
Since $\Omega \!\gg\! J,J',D$, we can capture these observations within a Floquet-Magnus high frequency
expansion. Given
the linear quadrupole-phonon coupling, we model the impact of the phonon drive on the spins directly via 
an effective `quadrupolar drive' $H^{\rm sp}_{\rm drive} = -A_{\rm eff} \sum_\br (T_{xz}(\br) \cos(\Omega t) + T_{yz}(\br) \sin(\Omega t))$, 
where $A_{\rm eff} = \lambda Q_0$ is the effective field, $Q_0$ is the steady state phonon amplitude (which
depends on the drive strength $A$) and $\lambda$ is the spin-phonon coupling constant. 
Since it is useful, we list the full set of spin-1 multipoles and their commutation relations in Appendix \ref{app:generators}.
% We understand the steady state physics of the full lattice model using
% a Floquet-Magnus expansion in the lab-frame, justified by $\hbar\Omega \gg J,J',D$ (see SM \cite{suppmat} for details).
% Explicitly, since the phonons couple linearly to the quadrupoles, we replace the phonon drive by
% $H^{\rm sp}_{\rm drive} = -A_{\rm eff} \sum_\br (T_{xz} \cos(\Omega t) + T_{yz} \sin(\Omega t))$, an effective quadrupolar drive.
% Here, the effective field $A_{\rm eff} = \lambda Q_0$ with $Q_0$ being the driven phonon amplitude (which
% depends on the original drive strength $A$) and $\lambda$ being the spin-phonon coupling constant. 
We Fourier decompose the drive Hamiltonian (see appendix \ref{app:Floquet-two-phonon-drive})
\begin{eqnarray}
    \!\!\!\! H^{\rm sp}_{\rm drive,0}\!&=&\!0; ~H^{\rm sp}_{\rm drive,\pm}\! = -\frac{\lambda Q_0}{2} \! \sum_\br (T_{xz}(\br) \mp i T_{yz}(\br))
\end{eqnarray}
A second-order Magnus expansion yields an effective Floquet Hamiltonian
\begin{eqnarray}
   \!\!\!\! H^{\rm sp,eff}_{\rm drive} \! &=& \frac{1}{\Omega} [H^{\rm sp}_{\rm drive,+}, H^{\rm sp}_{\rm drive,-}] \!= -\frac{\lambda^2 Q_0^2}{2\Omega} \sum_\br S_{z}(\br),
\end{eqnarray}
% In this way, the chiral 
% driving of the ${\cal E}_g$ phonon modes gives rise to an effective magnetic field in the $S_z$ direction. 
The phonon-induced effective magnetic field $(A_{\rm eff}^2/2\Omega)$ along $S_z$ inferred from this
Magnus expansion explains the enhanced magnetization at larger drive amplitude $A$. 
In the ordered AFM, this effective field
leads to a canting of the ordered moment, while in the QPM it leads to a weak uniform as well as staggered
magnetization (see appendix \ref{app:Floquet-two-phonon-drive}).
%, i.e. a non-equilibrium QPM to canted-AFM transition.
% \cite{suppmat}.

Previous work on ${\rm{CeCl_3}}$, a dipolar pseudospin-$1/2$ paramagnet, has proposed that a driven
chiral phonon can produce an effective magnetic field from coupling of the spin dipole moment 
to the chiral phonon angular momentum,
% \cite{juraschek2022giant},
which was studied using a mean-field rate equation \cite{juraschek2022giant}.
A microscopic theory of the inverse effect, a ``Zeeman splitting'' of chiral phonon 
modes by an applied magnetic field \cite{Intro_ChiralPhonon_chaudhary2023giant}, identified a mechanism where the phonon
couples a pair of spin-orbit split Kramers doublets in ${\rm{CeCl_3}}$. Our work proposes a distinct mechanism 
for phonon-driven magnetization, relying on the linear coupling of quadrupolar moments to the phonon coordinates,
and {our simulations go beyond mean field theory in
capturing the transient dynamics and spatio-temporal correlations on the lattice}.

% While previous work has shown that a magnetic field in systems with strong spin-orbital coupling
% can lead to a large ``Zeeman splitting''
% of chiral phonon modes \cite{Intro_ChiralPhonon_chaudhary2023giant}. This work explores a similar, and yet distinct effect, to the one discussed here where, in the paramagnetic ${\rm{CeCl_3}}$, an applied magnetic field was found to split degenerate chiral phonon modes. While another work, Ref.\cite{juraschek2022giant}, finds a similar effect in ${\rm{CeCl_3}}$ we note some important distinctions. Namely, Ref.\cite{juraschek2022giant} couples the local dipole moment of spins to the angular momentum of \emph{IR}-active phonon modes whereas we couple local quadrupolar moments linearly to the coordinates of Raman-active phonons. Further, Ref.\cite{juraschek2022giant} investigates this effect using a rate equation approach, while we here model the full semi-classical dynamics of the lattice coupled to phonons.

% We next study the full driven-dissipative 
% spin-phonon system on the lattice via $SU(3)$ dynamics simulation with a time-dependent chiral two-phonon drive.
% Fig.~\ref{fig:twoPhononDrive} reveals a phonon-induced uniform $S_z$ magnetization which settles to
% a near steady state after initial oscillatory transients, for various drive strengths. 

\section{Non-chiral single-mode drive}
% \section{Phonon-induced antiferromagnetic order}
% {\it Non-chiral single-mode drive. ---} 
We next investigate the effect 
of driving a single phonon mode, $Q_{xz}$ or $Q_{yz}$, on the spin model.
We model the $Q_{xz}$ drive via the Hamiltonian
\begin{equation}
H^{\rm{ph}}_{\rm{drive}}=-A \sum_\br \cos(\Omega t)Q_{xz}(\br),
\label{eq:1phononDrive}
\end{equation}
We show that this
results in the generation of AFM order starting from the QPM phase, signifying a 
non-equilibrium phase transition, and uniaxial spin anisotropy 
which leads to $O(2) \to Z_2$ symmetry lowering.

\begin{figure}[t]
\centering
\includegraphics[width=0.48\textwidth]{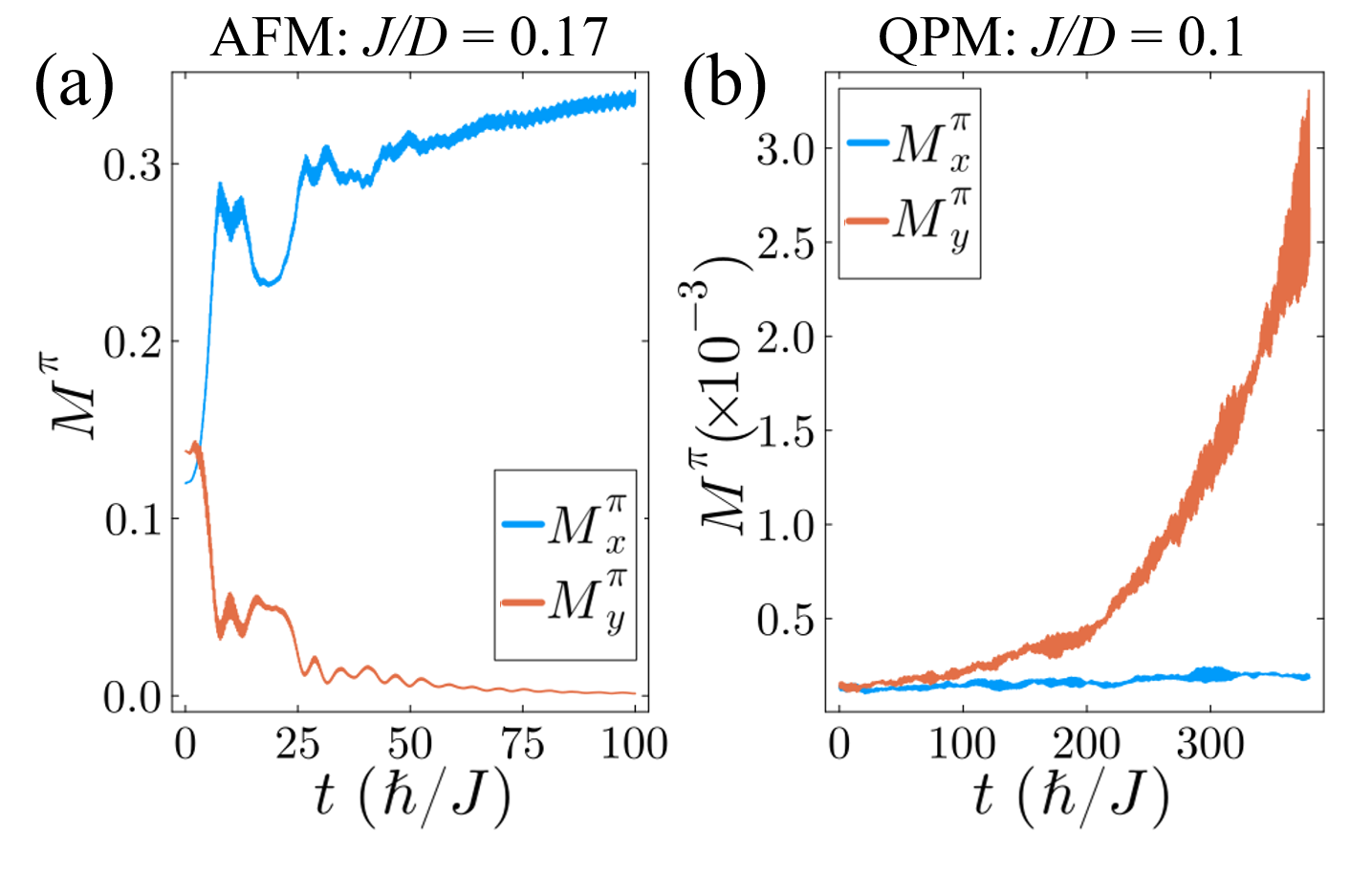}
\caption{Time evolution of in-plane components of the staggered magnetization $\vec M^{\pi}$, corresponding to wavevector $\bQ=(\pi,\pi,\pi)$, 
as a function of time for $T/J=0.2$ (averaged over 200 configurations) in (a) the AFM phase with $J/D=0.17$ and (b) the QPM phase with $J/D=0.1$. 
In the AFM, the staggered magnetization is strengthened along $S_x$ (i.e., $M^\pi_x$ is enhanced) due to a Floquet-drive induced Ising 
exchange anisotropy, while the QPM has enhanced $M^\pi_y$
due to Floquet drive induced quadrupole term $\sim\!T_{x^2-y^2}$.}
%which seems to nearly reach a steady state, indicating a possibly stable nonequilibrium phase.} 
\label{fig:AFMmag1}
\end{figure}

% where $A$ is the drive amplitude, $\Omega$ is resonant with the natural phonon frequency and $Q_{xz}(\br)$ is the local Einstein, ${\cal E}_g$ phonon. We again include a weak dissipation in the phonon sector in order to allow them to reach a steady state. 
Fig.~\ref{fig:AFMmag1}(a) shows the time evolution, up to $t \!\sim\! 100\hbar/J$,
of the staggered magnetization components $(M^\pi_x,M^\pi_y,M^\pi_z)$ averaged over $\sim\! 200$  
initial equilibrium AFM configurations.
Here, $M^\pi_i = |\sum_\br (-1)^\br \langle S_i(\br)\rangle|$ are components of the dipolar 
AFM order parameter. {We see that the XY symmetry of the equilibrium AFM leads to $M_x^\pi=M_y^\pi$ 
at initial time (with the small difference arising from averaging over a finite number of configurations)}; 
however, in the driven system $t>0$, $M_x^\pi(t)$ increases with time while $M_y^\pi(t)$
decreases and vanishes for $t \!\gtrsim\! 75/J$, so the $Q_{xz}$ drive generates an Ising anisotropy favoring $S_x$ AFM ordering.

Fig. \ref{fig:AFMmag1}(b) contrasts this with the time evolution starting from the equilibrium QPM phase. In this case, the initial
$\vec M^\pi = 0$; the observed small nonzero value at $t=0$ is due to finite size effects. When the drive is turned on, $t > 0$,
we find that the drive generates a small nonzero $M^\pi_y(t)$ while $M^\pi_x(t)$ stays
nearly zero. The system thus undergoes a QPM to Ising-AFM transition with a distinct $S_y$ Ising anisotropy unlike
the previous case.

% we see an $SO(2)$ symmetry breaking in both phases, however, the direction of the induced easy-axis AFM order depends on the initial state. In Fig. \ref{fig:AFMmag1} (a), we observe that the the staggered magnetization in $S_x$ increases over time, while that in $S_y$ decreases and reaches zero at $\sim 75/J$. In contrast, we find the opposite effect in the QPM phase, where the staggered magnetization in $S_y$ increases, while there is virtually no induced order in $S_x$ (Fig. \ref{fig:AFMmag1}(b)).

% The results of these simulations are shown in Fig. \ref{fig:AFMmag1}. In both QPM and AFM phases, the overall antiferromagnetic order increases over time. Furthermore, in (a), we clearly observe $SO(2)$ symmetry-breaking in that the AFM order in $S_x$ increases over time, while that in $S_y$ decreases. Remarkably, we find the opposite effect in the QPM phase (Fig. \ref{fig:AFMmag1}(b)).

To interpret the results seen in the full lattice model, we approximate the phonon drive as an effective drive acting directly on the 
quadrupole degrees of freedom, 
$H^{\rm sp}_{\rm{drive}}=-A_{\rm{eff}} \sum_\br \cos(\Omega t)T_{xz}(\br),$
where $A_{\rm{eff}} = \lambda Q_0$ with the phonon amplitude $Q_0$. 
{While the time-averaged
drive Hamiltonian does not break any symmetries}, a second order
Floquet-Magnus expansion (see appendix \ref{app:Floquet-one-phonon-drive}) yields an effective Hamiltonian
\begin{eqnarray}
\label{eq:magnusSpin1AFM}
   \! H^{\rm sp, eff}_{\rm drive} \!\! &=& \!\! \sum_{\langle \br\br'\rangle} \!\!\! J_{\br\br'} (1 \!-\! \frac{A_{\rm eff}^2}{2\Omega^2}) 
   [\bS (\br) \!\cdot\! \bS(\br')
   \!+\! (g\!-\! 1) S_y(\br)S_y(\br')] \nonumber\\
    \!\! &+& \! \Gamma \!\! \sum_{\langle \br\br'\rangle} \! (T_{xy}(\br) T_{xy}(\br')\!+\!T_{yz}(\br) T_{yz}(\br') \!+\! 
   T_{e}(\br) T_{e}(\br'))\nonumber\\
    &+&\! D (1 \!-\! 3 \frac{A_{\rm eff}^2}{4 \Omega^2})\! \sum_{\br}  S_z^2(\br) \!+\!
   \Delta \! \sum_\br T_{x^2-y^2}(\br)
\end{eqnarray}
where $T_e \!\equiv\! \sqrt{3} T_{z^2}\!-\!T_{x^2\!-\!y^2}$, $T_{z^2} \!=\! \frac{1}{\sqrt{3}}(3 S_z^2 \!-\! \bS^2$), $T_{x^2\!-\!y^2} \!=\! S_x^2\!-\!S_y^2$
and the newly generated couplings have strengths
\begin{equation}
    \Gamma \!=\! J \frac{A_{\rm eff}^2}{2\Omega^2};~ g \!=\! \frac{(2 \Omega^2 \!-\! 4 A_{\rm eff}^2)}{(2 \Omega^2\!-\!A_{\rm eff}^2)};~
    \Delta\!=\!D \frac{A_{\rm eff}^2}{4 \Omega^2}.
\end{equation}
We see that the phonon drive: (i) renormalizes both the Heisenberg spin exchange $J_{\br\br'}$
and the single-ion anisotropy $D$, (ii) generates Ising anisotropy $g \!<\! 1$
which leads to symmetry lowering $O(2) \to Z_2$, and
(iii) generates distinct quadrupolar exchange $\Gamma$ and
single-ion anisotropy term $\Delta$, {illustrating that the driven phonons in multipolar magnets
can be used to engineer unconventional effective spin Hamiltonians.}

A mean-field phase diagram of the effective Hamiltonian in Eq.\ref{eq:magnusSpin1AFM}, 
discussed in appendix \ref{app:Floquet-one-phonon-drive}, partially
explains two of our significant observations from the simulations. First, it shows that the QPM-AFM 
critical point shifts to smaller $J/D$ with increasing $A_{\rm eff}$; we trace this to the fact that 
the effective Hamiltonian has different renormalization factors for $J$ and $D$, thus enabling 
one to modify the $J/D$ ratio and drive the QPM to AFM transition. Second, the effective Hamiltonian 
breaks the $O(2)$ symmetry in the $S_x$-$S_y$ plane, since the drive also
differently renormalizes the $J_x/D$ and $J_y/D$ ratio in a manner which favors $S_x$-AFM when the dipolar magnetization
is significant as in the AFM phase. On the other hand, if we start in the
QPM phase, the dipolar terms in the Hamiltonian are less significant; the flipped Ising anisotropy in this case,
which favors $S_y$-AFM, can be traced back to the quadrupolar terms tied to $T_{x^2-y^2}$ with prefactors $\Gamma,\Delta$ in the
Floquet-Magnus effective Hamiltonian.

\section{Phonon-driven dynamical structure factor}
{
% {\it Phonon-driven dynamical structure factor.---} 
%We finally turn to the dynamical spin structure factor ${\cal S}(\bq,\omega)$ 
%of the spin-phonon coupled system which captures its full spatiotemporal spin correlations.
The dynamical response of equilibrium spin-$1$ magnets can be captured using
$SU(3)$ simulations for the spin structure factor
${\cal S}(\bq,\omega)$
\cite{do2021decay,do2023understanding}. Here, we focus on the previously
unexplored impact of phonons, especially
non-equilibrium driven phonons which lead to a gapped Nambu-Goldstone 
mode in the AFM and create higher Floquet spin excitation bands (see appendix \ref{app:DSSFcalc}).

\begin{figure}[t]
\centering
\includegraphics[width=0.48\textwidth]{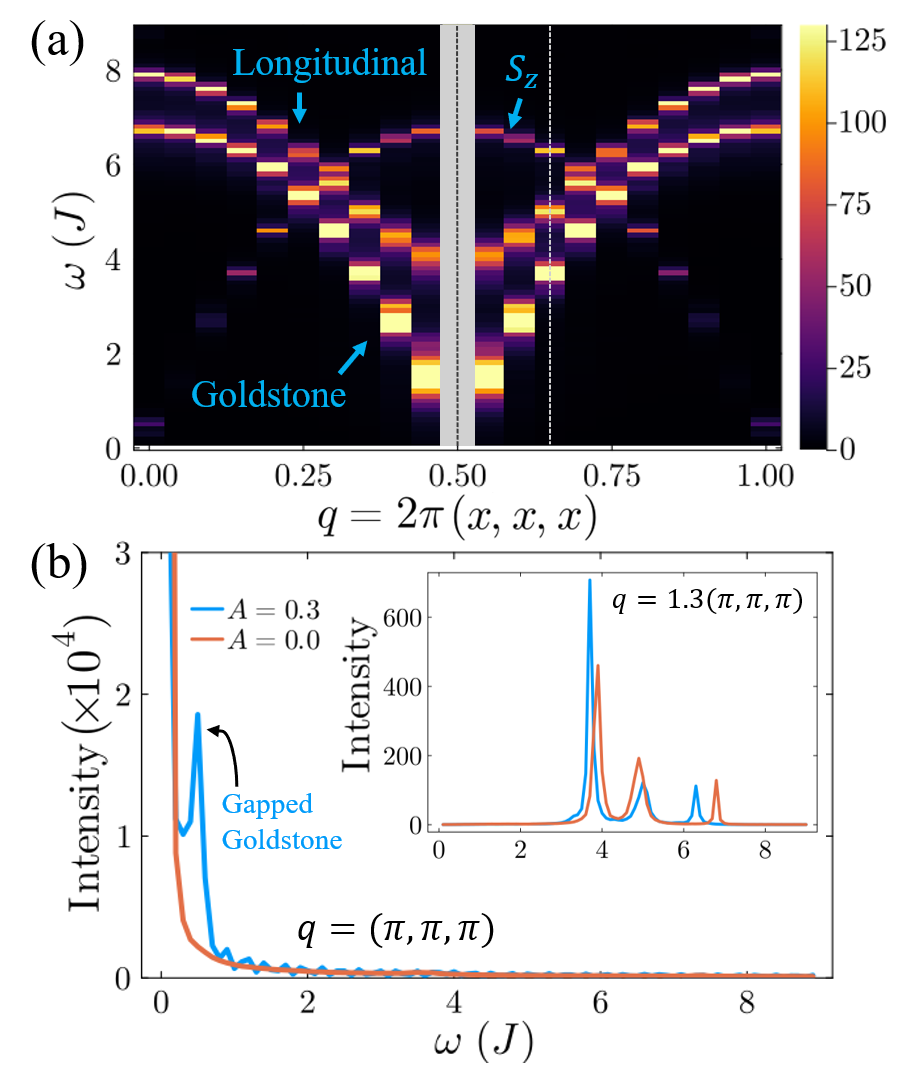}
\caption{(a) Dynamical spin structure factor $\mathcal{S}_{\rm dr}(\bq,\omega)$ with $Q_{xz}$-phonon drive in the 
N\'eel AFM revealing two transverse
modes, `$S_z$' and `Goldstone', corresponding to out-of-plane and in-plane fluctuations, and a
longitudinal mode due to quadrupolar fluctuations. We omit
$\bq=(\pi,\pi,\pi)$ due to intense diffuse scattering from the AFM order, but we show this as a cut in 
panel (b).
%The computed $\mathcal{S}_{\rm dr}(\bq,\omega)$ closely resembles the
%result $\mathcal{S}_{\rm eq}(\bq,\omega)$ in the equilibrium ordered phase. 
(b) Cut through $\mathcal{S}_{\rm dr}(\bq,\omega)$ (driven, blue: $A\!=\! 0.3$) and $\mathcal{S}_{\rm eq}(\bq,\omega)$ (undriven, orange: $A\!=\! 0.0$)
at fixed $\bq=(\pi,\pi,\pi)$. The emergent peak at $\omega\! \sim\! 0.4J$ in the driven spectrum
is the Floquet induced gapped Nambu-Goldstone mode. Inset: Cut at
$\bq\!=\!1.3 (\pi,\pi,\pi)$, with three sharp modes, and 
compared with undriven case to show mode energy renormalization. Cuts correspond to dashed lines in panel (a).}
\label{fig:spin1flavorwave}
\end{figure}

Fig.~\ref{fig:spin1flavorwave}(a) shows the low energy spin dynamical structure factor ${\cal S}_{\rm dr}(\bq,\omega)$ 
for the driven phonon system in the AFM phase at $J/D\!=\! 0.17$ and $T/J\!=\! 0.2$. We find a spectrum
with two highly dispersive transverse modes (labelled `Goldstone' and `$S_z$') and a less dispersive longitudinal mode.
{While this spectrum qualitatively resembles previous results obtained in the absence of phonons
\cite{do2021decay,do2023understanding}, there are important quantitative renormalizations  
due to both spin-phonon coupling and the 
non-equilibrium drive (see appendix \ref{app:DSSFcalc}); the impact of phonon drive is clearly visible in cuts through the 
spectrum at fixed $\bq$ as shown in Fig.~\ref{fig:spin1flavorwave}(b) inset.
Moreover, we observe a sharp qualitative
distinction between the undriven and driven phonon dispersions in the Nambu-Goldstone mode at 
$\bq\!=\!(\pi,\pi,\pi)$. While the undriven AFM has $O(2)$ symmetry and a gapless Nambu-Goldstone mode, the 
effective Floquet Hamiltonian in the driven AFM 
has only uniaxial Ising symmetry as discussed above, leading to a {\em gapped} Nambu-Goldstone 
mode, seen in Fig.~\ref{fig:spin1flavorwave}(b) as a sharp emergent peak at $\omega \! \approx \!0.4 J$. This peak
is absent for zero phonon-drive as seen from Fig.~\ref{fig:spin1flavorwave}(b), and its gap
is consistent with that deduced by fitting to the mode dispersion away 
from the ordering wavevector (see appendix \ref{app:DSSFcalc}). Massive Nambu-Goldstone modes
with tunable gap induced by explicit symmetry breaking have also
been proposed recently in stroboscopic Floquet systems \cite{hou2024floquetengineeredemergentmassivenambugoldstone}.}
Fig.~\ref{fig:spin1flavorwave}(b) inset also
compares the driven and undriven mode energies at $\bq=1.3(\pi,\pi,\pi)$ showing that the
mode energies are renormalized by the phonon drive when compared with the equilibrium undriven ($A=0$)
phonon case ${\cal S}_{\rm eq}(\bq,\omega)$.
The phonon drive also induces phonon-dressed Floquet spin wave bands; see appendix \ref{app:DSSFcalc} for details.

\section{Discussion}
% {\it Discussion.---}
{We have developed a technique to calculate spin-phonon dynamics in higher-spin multipolar Mott insulators featuring 
quadrupole moments linearly coupled to driven dissipative Einstein phonons, and uncovered a wide range of remarkable
phenomena which could potentially be observed in a candidate material}. Our results on the generation of magnetization or the
canted AFM with chiral phonons could be explored using time-resolved Kerr rotation after using a Raman process to
pump phonons \cite{Padmanabhan_NatComms2022}. If inversion is weakly broken, it may allow phonons to be
driven directly using the electric field of light \cite{nova2017effective,afanasiev2021ultrafast,Intro_ShastrySutherland_Ruegg_PRB2023}. 
The phonon-drive induced renormalization of the mode energies
could be detected at $\bq=0$ using THz spectroscopy. Finally,
Floquet states of electronic matter have been fruitfully explored using time and angle resolved photoemission
spectroscopy in topological insulators and graphene \cite{floquet_trARPES_gedik_2013,floquet_trARPES_mathias_2024,floquet_trARPES_gedik_2024}.
An analogous next generation experiment to study the full non-equilibrium ${\cal S}_{\rm dr}(\bq,\omega)$ discussed in our work
would be inelastic neutron scattering in the presence of driven phonons \cite{Pump_neutron_2024}.
Our work presents an exciting vehicle to study the 
dynamics of multipolar systems including important lattice degrees of freedom, 
and suggests new avenues in non-equilibrium multipolar magnetism.

\noindent {\bf Acknowledgements:}
This research was funded by an NSERC Discovery Grant (AP),
an Ontario Graduate Scholarship (KH), and an NSERC CGS-D fellowship (RS).
We thank Emily Zhang, William Bateman-Hemphill, and Cristian Batista for helpful discussions.
Numerical computations were performed on the Niagara supercomputer at the SciNet HPC 
Consortium and the Digital Research Alliance of Canada.

\bibliography{main}

% \newpage
\onecolumngrid
\appendix

\section{Details of SU(N) spin-phonon MC}
\label{app:MCdetails}

In this section, we provide a detailed discussion of the $SU(N)$ Monte Carlo (MC) and molecular dynamics (MD) simulations used in this work. 
Previous work in Refs.\cite{remund2022semi,dahlbom2022geometric,do2023understanding} has developed the $SU(N)$ framework for equilibrium and dynamical simulation of $S>1/2$ magnets. 
Explicitly, these simulations overcome shortcomings of conventional vector MC, which is grounded in an $SU(2)$ representation of spins, where each point on the Bloch sphere is mapped to a three-dimensional vector representing the dipolar expectation values of an individual spin. 
Metropolis MC updates proposed within this framework are then represented by $SU(2)$ rotations on this sphere. 
It is important to note, however, that there is \emph{no} such rotation which will transform a dipolar state into a multipolar one. 
Alternatively, $SU(N)$ simulations, where $N=2S+1$ for a spin-$S$ system follow a similar principle, but are generalized to treat local dipolar and multipolar moments on equal footing. 
To do this, we begin by representing local spins as \emph{wavefunctions}, $|\psi_\br\rangle$, $2S+1$ dimensional complex vectors constrained by normalization and the irrelevance of phase, given by Eq. \ref{eq:wavefunction}. 
One may then construct a semi-classical simulation where the global wavefunction is $|\Psi\rangle=\otimes_\br |\psi_\br\rangle$, ignoring entanglement between spins while capturing the on-site quantum mechanics \cite{remund2022semi,dahlbom2022geometric}. 
Using these local wavefunctions one is able to track the expectation values of all $N^2-1$ generators of the representation from which all dipolar and \emph{multipolar} expectation values may be constructed. 
%For $S=1/2$, a choice of $N=2$ and the three generators of the representation as the Pauli matrices reveals the equivalence to conventional MC simulations. As an example, local wavefunctions for the spin-$1/2$ and spin-$1$ cases are shown below:
%

For the spins, we implement a Metropolis update scheme, wherein a site from the lattice is chosen at random and a new state $|\psi^{\rm{new}}_\br \rangle$ is proposed as a new random $N$ dimensional vector of complex numbers. 
Next, the change in energy $\Delta E$ is calculated using the spin Hamiltonian $E_{\rm sp} = \langle H_{\rm sp} \rangle$ (and the spin-phonon coupled Hamiltonian $E_{\rm sp-ph} = \langle H_{\rm{sp-ph}} \rangle$, discussed below) and updates are then either accepted or rejected with probability $\exp{-\Delta E/T}$ if the change in energy is greater than zero and with probability of unity if the change in energy is less than zero.
% Next, the expectation values of the spin, $H_{\rm{sp}}$, and spin-phonon coupled, $H_{\rm{sp-ph}}$, are calculated. Updates are then either accepted or rejected with probability $\exp{(-(\langle H_{\rm{sp}}\rangle + \langle H_{\rm{sp-ph}}\rangle)/T}$ if the change in energy is greater than zero and with probability of unity if the change in energy is less than zero.

Our work extends these $SU(N)$ simulations, adding Einstein phonons to this algorithm. 
Explicitly, each phonon is represented by two, classical, degrees of freedom; a coordinate and a momentum. 
In practice for the purposes of Monte Carlo simulations, we integrate out the momentum degrees of freedom as they do not couple to the spins. 
Thus, in an $SU(N)$-phonon MC simulation we are left with an additional $\alpha N_{\rm{spins}}$ extra degrees of freedom, where $\alpha$ is the number of phonons included, as compared to previous $SU(N)$ spin-only simulations.

In practice, $SU(N)$ simulations including phonons proceed along similar lines to those without. 
To begin, although the phonon coordinate is inherently bounded by the quadratic potential in which it sits, we select a $Q_{\rm{max}}\sim10\%$ of the lattice constant by which to bound the phonon coordinate. 
Explicitly, at the beginning of simulations, we initialize all $\alpha  N_{\rm{spins}}$ from a uniform distribution ranging from $[-Q_{\rm{max}},Q_{\rm{max}}]$. 
Simulations then proceed along the lines  of conventional Metropolis update algorithms, wherein, we choose one site from the lattice and propose an update to all phonons on the corresponding lattice site. 
The update of each phonon on the lattice site is then checked individually and accepted if it reduces the energy of the system and accepted with a probability $\exp{-\Delta E/T}$ if it raises the energy. 
Throughout the simulation we update both the spins and phonons in tandem, performing alternating `sweeps' (one update per spin/phonon) of the lattice, sweeping through first the spins and subsequently the phonons. 
We track the energy of the spins, phonons and spin-phonon coupling separately throughout the simulation so as to ensure that measurement sweeps are only performed once both spins and phonons are thermalized.

%  \textcolor{red}{moved this para from main text - need to fit it in}.
%  In practice, we pick phonon coordinates from a bounded box 
%  $(-\!Q_{\rm max}\!,\!Q_{\rm max})$; see Supplemental Material (SM) \cite{suppmat} for details.
% To simulate dynamics at any temperature $T$, we start from an equilibriated MC configuration for spin wavefunctions and phonon coordinates,
%  and pick initial phonon momenta from a Boltzmann distribution at temperature $T$. This initial configuration is then time-evolved 
%  according to the Schr\"odinger equation for the spin state and Newton's laws for $\{Q,P\}$ to compute spatiotemporal correlations. 
%  We finally average all observables over a large number $\sim 200$-$300$ initial MC configurations. Our implementation of the time-integration
%  uses the Schrodinger midpoint method discussed in Ref.\cite{dahlbom2022geometric}, with an additional step where we 
%  update both phonon  coordinates and momenta via a matching midpoint scheme, while satisfying energy 
%  conservation and all multipolar operator trace constraints; see SM \cite{suppmat} for details.

In our work, we additionally implement a parallel tempering algorithm as discussed in Ref. \cite{hukushima1996exchange}. 
However, we necessarily make the addition that, in order to exchange configurations at nearby temperature points, both the spin energies and the phonon energies of the two configurations must each be similar \emph{individually}. 
Exchange of configurations based on total energy alone proves problematic when calculating variance-dependent quantities.

\section{Details of SU(N) spin-phonon MD}
\label{app:MDdetails}
We next move to a discussion of the dynamics simulations used in this work. Perhaps the most common technique used to simulate the dynamics of spins is through the famous Landau-Lifshitz (LL) equations of motion (EOM). These equations describe the dynamics of spin operators taken in the classical limit using SU(2) spins and thus, similarly to conventional MC, do not permit the dynamical evolution of higher order moments. Therefore, higher spin systems need to be modeled using $SU(N)$ coherent states which take all multipolar operators into account. Although previous work has shown the ability to evolve spin operators directly using classical equations of motion, we opt to evolve the wavefunctions directly using the Schrodinger Midpoint method outlined in Ref. \cite{dahlbom2022geometric}. The advantage of this method is that such a simulation need only work with $2N$ real parameters of the complex wavefunctions, rather than $N^2-1$ local operator expectation values. 

Using such a construction, local wavefunctions at each site are evolved in the Schrodinger picture using a mean-field Hamiltonian
\begin{equation}
\label{eq:schrod_eq}
    \frac{d}{dt}| \psi_\br (t) \rangle = - i\mathcal{H} |\psi_\br (t) \rangle
\end{equation}
Where $\mathcal{H}$ is given by the local mean-field Hamiltonian. We use the midpoint state
$$
|\tilde{\psi_\br} \rangle = \frac{| \psi_\br (t') \rangle + |\psi_\br (t) \rangle}{2}
$$
to iteratively evolve the state through a time $\delta t = t' - t$. For each site in the lattice, we begin by setting the final state equal to the initial state,
$
| \psi_\br (t') \rangle = |\psi_\br (t) \rangle
$
% This implies that the midpoint state is also equal to the initial state. 
We then update the final state $| \psi_\br (t') \rangle$ using Eq. \ref{eq:schrod_eq}, and repeat this process, iteratively updating the state at $t'$ until convergence. In implementing this Schrodinger midpoint solver, we were able to reproduce energy conservation similar to that reported in Ref. \cite{dahlbom2022geometric}.

For this work, we extended this solver to include phonon coordinates $Q_\alpha$ and momenta $P_\alpha$, updated using a similar midpoint scheme. Explicitly, we begin these simulations using a set of phonon coordinates, $Q_\alpha(\br)$, from a thermalized MC configuration as initial conditions. However, since we do not simulate the phonon momenta in the MC simulations, we instead begin MD simulations by drawing the phonon momenta from a Boltzmann distribution at the appropriate temperature. Using Newton's equations of motion, we then evolve the phonon coordinates and momenta using the mean-field Hamiltonian by
\begin{eqnarray}
\label{phonon_EOM}
    \frac{dQ_\alpha}{dt}=P_\alpha/M_\alpha; \quad
    \frac{dP_\alpha}{dt}=-M_\alpha\Omega_\alpha^2Q_\alpha -\lambda_\alpha \langle T_\alpha \rangle
\end{eqnarray}
where $M_\alpha$ is the mass of the phonon mode and $\Omega_\alpha$ is the natural frequency of the phonon. Note that these equations resemble the classical harmonic oscillator with an additional term given by the spin-phonon coupling $H_{\rm sp-ph}$. We evolve the phonons in a similar midpoint scheme to the spins: starting by setting the final position $Q_\alpha(t')$ and momentum $P_\alpha(t')$ to the initial position and momentum, and solve the midpoint coordinate by
\begin{eqnarray}
\label{phonon_midoint}
    \tilde{Q}_\alpha = \frac{Q_\alpha(t') + Q_\alpha(t)}{2};\quad \tilde{P}_\alpha = \frac{P_\alpha(t') + P_\alpha(t)}{2}.
\end{eqnarray}
We update the final values using the equations of motion given in Eq. \ref{phonon_EOM}, and continue until the momentum and position coordinates converge. Evolving the spin wavefunctions and phonons is done simultaneously, meaning that for each iteration of the midpoint method, all spin wavefunctions, phonon coordinates and phonon momenta are updated. We show plots of the energy drift in the Sec. \ref{sec:SUN-MD-tests}, in which we get considerable energy conservation for the entire spin-phonon coupled system. In many of the simulations conducted for this work, however, we also include terms which do not conserve energy, explicitly both a damping and driving term in the phonon equations of motion. We represent this damping in the form of a $-\eta P_\alpha$ in the momentum equation of motion, where $\eta$ is the damping constant and is set to 0.1 in all simulations to ensure the reaching of a steady state. Further, we include any driving terms for the phonons, coupled to the phonon coordinate in the driving Hamiltonian, and evaluated at the midpoint time $\tilde{t}=\frac{t'+t}{2}$. Additionally, we find it important to note that, for calculations of dynamical observables, we set a rate $\epsilon$ at which we sample configurations of the system at time steps of $\Delta t = \epsilon \delta t$. Error in the averaged MD dynamics data in this work is not shown in the plots as the error (of order $\sim 0.03\%$) is not statistically significant and would not be visible.
\section{Numerical tests of SU(N) Monte Carlo and Molecular Dynamics of spins and phonons}
\label{app:tests}

\subsection{SU(N) spin-phonon MC}

As a test case for the SU(N) MC, we construct a nearest-neighbour spin-1 Heisenberg antiferromagnetic spin model coupled to two phonons on a cubic lattice. Explicitly, one may write:
$$
H=H_{\rm{sp}}+H_{\rm{ph}}+H_{\rm{sp-ph}}
$$
where, $H_{\rm{sp}}=J\sum_{\langle\br\br'\rangle}S(\br)\cdot S(\br')$, $H_{\rm{sp-ph}}=\lambda  \sum_\br\left(  T_{xz}(\br)Q_{xz}(\br)+T_{yz}(\br)Q_{yz}(\br) \right)$ and $H_{\rm{ph}}=\frac{1}{2} M\Omega^2\sum_\br(Q_{xz}(\br)^2+Q_{yz}(\br)^2) $ where we have set the masses, $M$ and frequencies $\Omega$ to be the same for both phonons. We run a parallel-tempered MC simulation with 80 temperature points for $10^6$ sweeps of a lattice of size $10^3$. The resulting heat capacity $C_v$ of this simulation is shown in Fig.\ref{fig:HCtest}, where we observe the expected, constant, contribution of the phonons $C_{\rm{ph}}=1$ from the two quadratic degrees of freedom (phonon displacements) in addition to the expected $C_{\rm{sp}}$ from such an antiferromagnetic model.

\begin{figure}[h]
\includegraphics[height=0.25\textwidth]{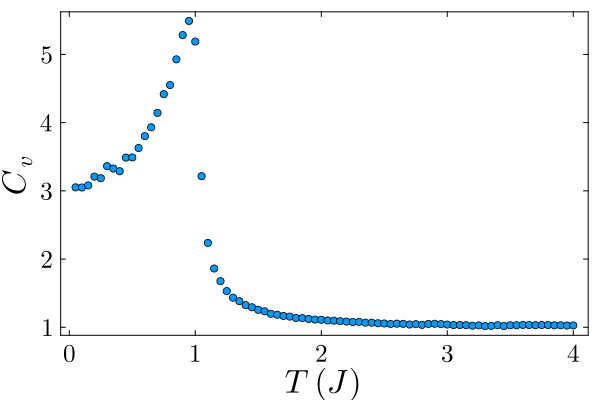}
\caption{Heat capacity for an antiferromagnetic $S=1$ Heisenberg model coupled to two phonons on the cubic lattice. We computed 80 temperature points, using parallel tempering and $10^6$ sweeps. We see a constant contribution of $C_{\rm{ph}}$ from the two quadratic phonons terms, in addition to the expected $C_{\rm{sp}}=2$ in the low-temperature limit arising from the dipolar and quadrupolar degrees of freedom.}
\label{fig:HCtest}
\end{figure}

\subsection{ SU(N) spin-phonon MD}
\label{sec:SUN-MD-tests}

\begin{figure*}[t]
\centering
\includegraphics[width=\textwidth]{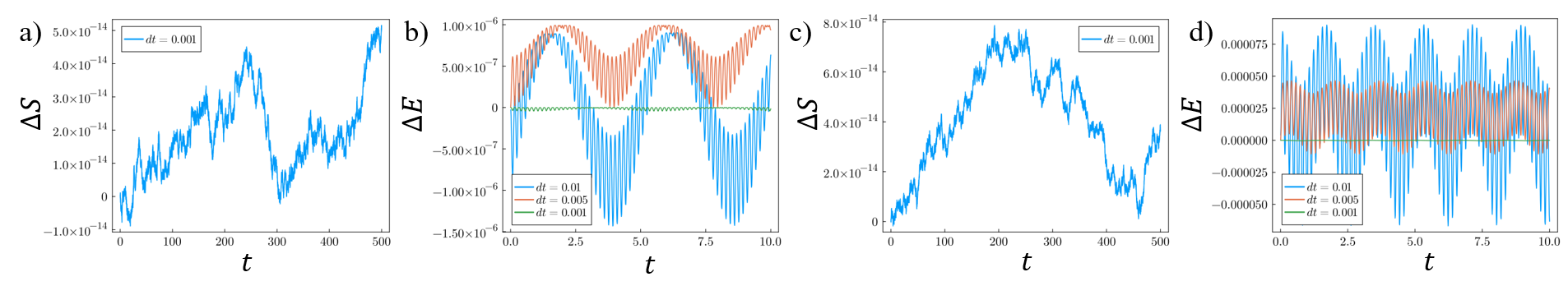}
\caption{Numerical checks for the spin-1 model (a),(b) and spin-3/2 model (c),(d), with a single spin in a magnetic field and a single phonon coupled to $Q_{xz}$. For $\delta t=0.001$, we show (a),(c) spin length conservation out to $t=500$. (b),(d) Energy conservation for three different $\delta t$ values, showing that the conservation is improved with smaller time step. We used values similar to those used in our numerical simulations.  
%, namely $g_1 \approx 80$, $g_2 \approx 4$ and $g_3 \approx 2.5$.
}
\label{fig:spinEnergyCons}
\end{figure*}

Phonons exist at an energy scale much higher than that of the spins, thus, coupling them to the spins leads to additional numerical constraints which are not present in the bare spin dynamics simulations. Explicitly, in order to avoid divergences in the phonon coordinate over long timescales, the time step $\delta t$ must be set such that it is much smaller than the period of the phonon modes. This requirement, along with the improved energy and spin length conservation at lower $\delta t$ (Fig. \ref{fig:spinEnergyCons}), has led us to use time steps of order $\sim 10^{-3}$ in order to simulate phonons of energies up to 80 meV. We have tested this method using a simple toy model, in the spin-1 and spin-3/2 cases, of a single spin with a single phonon mode sitting in a small magnetic field, whose dynamics are governed by
$$
H_{\rm sp} = -B_zS_z;\quad H_{\rm sp-ph} = \lambda Q_{xz}(\br)T_{xz}(\br)
$$
where $B_z$ is a small field in $z$, $Q_{xz}(\br)$ is the phonon coordinate and $T_{xz}(\br)$ is the $xz$ quadrupole. 
Here, we choose to write our Hamiltonian in units of the magnetic field $B_z = B$, thus, for the numerical test simulations, we select a phonon frequency of $\Omega_{xz}/B \approx 40$, and $\lambda/B = 2.5$.
In the spin-1 case, the spin is represented by $SU(3)$ coherent states and realizes $8$ spin moments ($3$ dipolar and $5$ quadrupolar). With a single phonon coupled, we get an initial configuration for this system using the phonon-coupled Monte Carlo at a temperature of $T=0.1$. We then run the SU(N)-phonon molecular dynamics without damping or driving the phonon, in order to show energy conservation for long times. Fig. \ref{fig:spinEnergyCons}(a),(b) shows the spin length and energy conservation for such an $SU(3)$ spin coupled to a single $Q_{xz}$ phonon. 

To further test these simulations, we run the same model for a spin-3/2 system. For this model, the spins are represented by $SU(4)$ coherent states and realize an additional 7 octupolar components. We show conservation of spin length and energy for this simulation in Fig. \ref{fig:spinEnergyCons}(c),(d) for a single $SU(4)$ spin coupled to a single $Q_{xz}$ phonon at the same energy. Note that, while we do not show the energy conservation up to long timescales, we find that there is no significant change in the energy up to $t=500$. We also find that the energy drift increases slowly at higher $\delta t$.

\section{Spin-1 multipoles, commutators and Gell-Mann matrices}
\label{app:generators}

In this section, we introduce a number of useful definitions for the spin-1 model described in this work. Using the dipolar
spin operators $S_x,S_y$ and $S_z$ which satisfy the usual $SU(2)$ algebra, we
define quadrupolar operators which are quadratic in the spin-1 operators and may be arranged into traceless tensors:
\begin{eqnarray}
    T_{xy} &=& S_xS_y + S_yS_x\\
    T_{xz} &=& S_xS_z + S_zS_x\\
    T_{yz} &=& S_yS_z + S_zS_y\\
    T_{x^2-y^2} &=& S_x^2 - S_y^2\\
    T_{z^2} &=& \frac{1}{\sqrt{3}}(3S_z^2 - 2)
\end{eqnarray}
We have set the normalization of all dipole and quadrupole operators ${\cal O}$ so they all have $\Tr({\cal O}^2)=2$.
The full set of commutation relations is given by

{\bf Dipole-Dipole:}
\begin{eqnarray}
\left[S_x,S_y\right] &=& i S_z ; ~~
\left[S_y,S_z\right] = i S_x ;~~
\left[S_z,S_x\right] = i S_y
\end{eqnarray}

{\bf Quadrupole-Quadrupole:}
\begin{eqnarray}
    \left[T_{xy},T_{xz}\right] &=& iS_x\quad
    \left[T_{xy},T_{yz}\right] = -iS_y \quad
    \left[T_{xy},T_{x^2-y^2}\right] = -2iS_z \quad
    \left[T_{xy},T_{z^2}\right] = 0 \nonumber \\
    \left[T_{xz},T_{yz}\right] &=& iS_z\quad
    \left[T_{xz},T_{x^2-y^2}\right] = iS_y\quad
    \left[T_{xz},T_{z^2}\right] = -\sqrt{3}iS_y \nonumber \\
    \left[T_{yz},T_{x^2-y^2}\right] &=& iS_x\quad
    \left[T_{yz},T_{z^2}\right] = \sqrt{3}iS_x \nonumber \\
    \left[T_{x^2-y^2},T_{z^2}\right] &=& 0,
\end{eqnarray}

{\bf Quadrupole-Dipole}:
\begin{eqnarray}
    \left[S_x,T_{xy}\right] \!&=&\! iT_{xz} \quad
    \left[S_x,T_{xz}\right] \!=\! -iT_{xy} \quad
    \left[S_x,T_{yz}\right] \!=\! i(\sqrt{3}T_{z^2}\!+\!T_{x^2-y^2}) \quad
    \left[S_x,T_{x^2-y^2}\right] \!=\! -iT_{yz} \quad
    \left[S_x,T_{z^2}\right] \!=\! -i\sqrt{3}T_{yz} \nonumber \\
    \left[S_y,T_{xy}\right] \!&=&\! -iT_{yz} \quad
    \left[S_y,T_{xz}\right] \!=\! -i(\sqrt{3}T_{z^2}\!-\! T_{x^2-y^2}) \quad
    \left[S_y,T_{yz}\right] \!=\! iT_{xy} \quad
    \left[S_y,T_{x^2-y^2}\right] \!=\! -iT_{xz} \quad
    \left[S_y,T_{z^2}\right] \!=\! i\sqrt{3}T_{xz} \nonumber \\
    \left[S_z,T_{xy}\right] &=& -iT_{x^2-y^2} \quad
    \left[S_z,T_{xz}\right] = iT_{yz} \quad
    \left[S_z,T_{yz}\right] = -iT_{xz} \quad
    \left[S_z,T_{x^2-y^2}\right] = iT_{xy}\quad
    \left[S_z,T_{z^2}\right] = 0.
\end{eqnarray}

While there are, naturally, a large number of individual commutation relations, it is important to note that there are a number of closed loops that we have made use of in the parts of this work concerning the spin-1 model. Specifically, one may note that the commutation relations of:
\begin{itemize}
\item $T_{xz}, T_{yz}$, and $S_z$
\item $S_{y}, T_{xz}$, and $\sqrt{3}T_{z^2}-T_{x^2-y^2}$,
\item $S_{x}, T_{yz} $, and $\sqrt{3}T_{z^2}+T_{x^2-y^2}$,
\end{itemize}
are each respectively self-contained.

We also present the generators used in the SU(3), $S=1$, simulations in this work. To this end, we use the Gell-Mann matrices:
\begin{align*}
\lambda_1 &= 
    \begin{pmatrix}
    0 & 1 & 0\\
    1 & 0 & 0\\
    0 & 0 & 0\\
    \end{pmatrix} &
    \lambda_2 &= 
    \begin{pmatrix}
    0 & -i & 0\\
    i & 0 & 0\\
    0 & 0 & 0\\
    \end{pmatrix} &
    \lambda_3 &= 
    \begin{pmatrix}
    1 & 0 & 0\\
    0 & -1 & 0\\
    0 & 0 & 0\\
    \end{pmatrix} &
    \lambda_4 &= 
    \begin{pmatrix}
    0 & 0 & 1\\
    0 & 0 & 0\\
    1 & 0 & 0\\
    \end{pmatrix} \\
    \lambda_5 &= 
    \begin{pmatrix}
    0 & 0 & -i\\
    0 & 0 & 0\\
    i & 0 & 0\\
    \end{pmatrix} &
    \lambda_6 &= 
    \begin{pmatrix}
    0 & 0 & 0\\
    0 & 0 & 1\\
    0 & 1 & 0\\
    \end{pmatrix} &
    \lambda_7 &= 
    \begin{pmatrix}
    0 & 0 & 0\\
    0 & 0 & -i\\
    0 & i & 0\\
    \end{pmatrix} &
    \lambda_8 &= \frac{1}{\sqrt{3}}
    \begin{pmatrix}
    1 & 0 & 0\\
    0 & 1 & 0\\
    0 & 0 & -2\\
    \end{pmatrix}
\end{align*}
Since we find it useful in our $SU(3)$ simulations \cite{do2023understanding,remund2022semi}, we also present the decomposition of spin-1 dipole and quadrupole operators into a linear combination of the Gell-Mann matrices:
\begin{align*}
    S_x &= \frac{1}{\sqrt{2}}(\lambda_4 + \lambda_6) &    
    S_y &= \frac{1}{\sqrt{2}}(\lambda_5 - \lambda_7) &
    S_z &= \lambda_3\\
    T_{xy} &= \lambda_2 &
    T_{xz} &= \frac{1}{\sqrt{2}}(\lambda_4 - \lambda_6) &
    T_{yz} &= \frac{1}{\sqrt{2}}(\lambda_5 + \lambda_7) \\
    T_{x^2-y^2} &= \lambda_1 &
    T_{z^2} &= \lambda_8.
\end{align*}

\section{Effective Hamiltonian and AFM order in the two-phonon drive}
\label{app:Floquet-two-phonon-drive}

In this section, we provide a description of the lab frame Floquet-Magnus expansion used in this work, specifically in the two-phonon drive. There has been significant recent theoretical and experimental interest in \emph{time-periodic} systems and Hamiltonians of which this work is but one example \cite{liu2018floquet,junk2020floquet,nuske2020floquet}. To describe such systems, it has become commonplace to use Floquet theory, decomposing the dynamics of the system into a product of unitary operators describing dynamics within the period of the system and longer time dynamics, extending over multiple periods \cite{eckardt2015high}. Although Floquet theory provides a neat description of time-periodic Hamiltonian systems, in practice, computation of the operators governing the dynamics of the system (the stroboscopic kick operator for short-time dynamics and the effective Hamiltonian for long-time dynamics) can be challenging, or impossible, to compute exactly. Thus, in systems where the drive frequency, $\Omega$, is significantly higher than other energy scales in the problem, one can employ an expansion in powers of $1/\Omega$ to calculate the effective Hamiltonian. 

To employ the Floquet-Magnus expansion, we begin with a decomposition of the Hamiltonian into its spectral components, $H_k$, where one may define:
$$
H(t)=\sum_{k=-\infty}^\infty H_k e^{ik\omega t}.
$$
Having calculated the relevant $H_k$, a computation of $H_{\rm{eff}}$, where $H_{\rm{eff}}=\sum_{j}H^{(j)}_{\rm{eff}}$ is possible. Explicitly, for this work, we computed $H_{\rm{eff}}$ to leading order, where: 
\begin{equation}
H_{\rm{eff}}\simeq H^{(1)}_{\rm{eff}}=\frac{1}{\Omega} \left[H_+,H_{-}\right]
\end{equation}
such that $k=+1$ corresponds to $H_+$ and $k=-1$ corresponds to $H_-$. In the main text, we compute the Magnus expansion on $H_{\rm drive}^{\rm sp}$,
\begin{eqnarray}
\label{eq:2siteMF}
   \!\!\!\! H^{(1)}_{\rm{eff}} \! &=& \frac{1}{\Omega} [H^{\rm sp}_{\rm drive,+}, H^{\rm sp}_{\rm drive,-}] \!= -\frac{\lambda^2 Q_0^2}{2\Omega} \sum_\br S_{z}(\br),
\end{eqnarray}
Thus, our effective Hamiltonian may be written as $H_{\rm{eff}}=H^{(0)}_{\rm{eff}}+ H^{(1)}_{\rm{eff}}$, where $H^{(0)}_{\rm{eff}}$ is the equilibrium Hamiltonian. To highlight the dynamics of the antiferromagnetic magnetization under resonant excitation by the two phonon drive, we construct a zero-temperature two-site mean-field phase diagram for this Hamiltonian, with results shown in Fig.\ref{fig:MFphaseDiagrams-2phonon}. We note a twofold effect as the resonant phonon amplitude increases: firstly, a downward renormalization of the critical point from its equilibrium value of $J/D\simeq 0.12$ (Fig.\ref{fig:MFphaseDiagrams-2phonon}(a)), and secondly, the ferromagnetic canting of the easy-plane AFM order (Fig.\ref{fig:MFphaseDiagrams-2phonon}(b)). The results from the full numerical model in the main text (Fig. \ref{fig:twoPhononDrive}) highlight the $S_z$ canting of the AFM order, and we additionally present results here highlighting the shifting of the critical point that is induced by the drive. Explicitly, in Fig.\ref{fig:AFMmag--2phonon} we show the total antiferromagnetic magnetization as a function of time to $t=380/J$ averaged over $200$ configurations undergoing a two-phonon drive with initial configurations in (a) the AFM phase and (b) the QPM phase. In Fig.\ref{fig:AFMmag--2phonon}(a), we  note an increase in the AFM order of $\sim 40\%$, which is qualitatively consistent with the shifting of the critical point seen in  Fig.\ref{fig:MFphaseDiagrams-2phonon}(a). In panel (b) we see a small increase in the AFM magnetization indicating that the two-phonon drive can also lead to a QPM-to-AFM transition, as indicated by the mean-field phase diagram (Fig.\ref{fig:MFphaseDiagrams-2phonon}(a)), but does not show it directly.
\begin{figure}[H]
\centering
\includegraphics[width=0.95\textwidth]{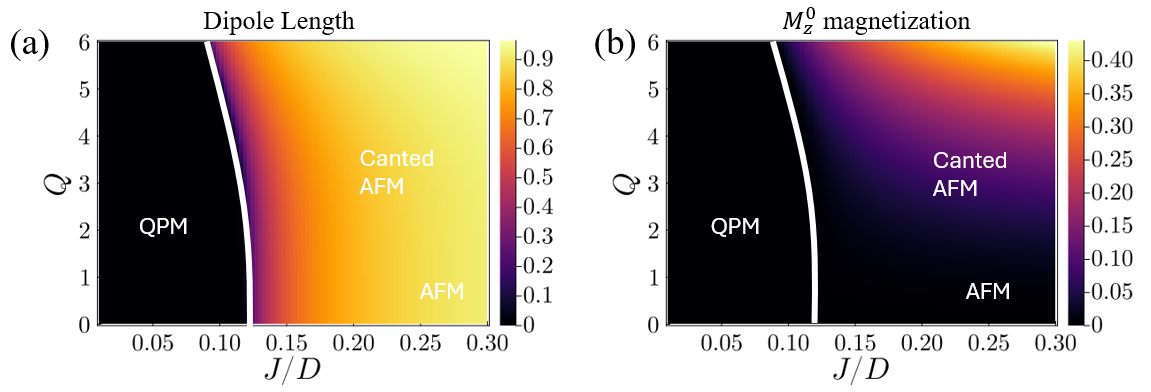}
\caption{Mean field phase diagram of the effective Hamiltonian, $H_{\rm{eff}}=H^{(0)}_{\rm{eff}}+H^{(1)}_{\rm{eff}}$, for the spin 1 model in a chiral two-phonon drive as a function of both $J/D$, the ratio of spin exchange to single-ion anisotropy in the static model, to phonon amplitude $Q$ which scales effective drive strength $A_{\rm{eff}}$. We highlight both the dipole length, (a), and magnitude of the $S_z$ dipole component, (b). With $Q=0$ we recover the static case, observing a quantum paramagnetic regime at small $J/D$ that transitions to an easy-plane $xy$ antiferromagnet with increasing $J/D$. With increasing phonon amplitude, we note a renormalization of the QCP away from its equilibrium value of $J/D\simeq 0.12$. Further, as shown in panel (b) we note a ferromagnetic canting of the AFM order away from the equilibrium easy XY-plane.}
\label{fig:MFphaseDiagrams-2phonon}
\end{figure}

\begin{figure}[H]
\centering
\includegraphics[width=0.95\textwidth]{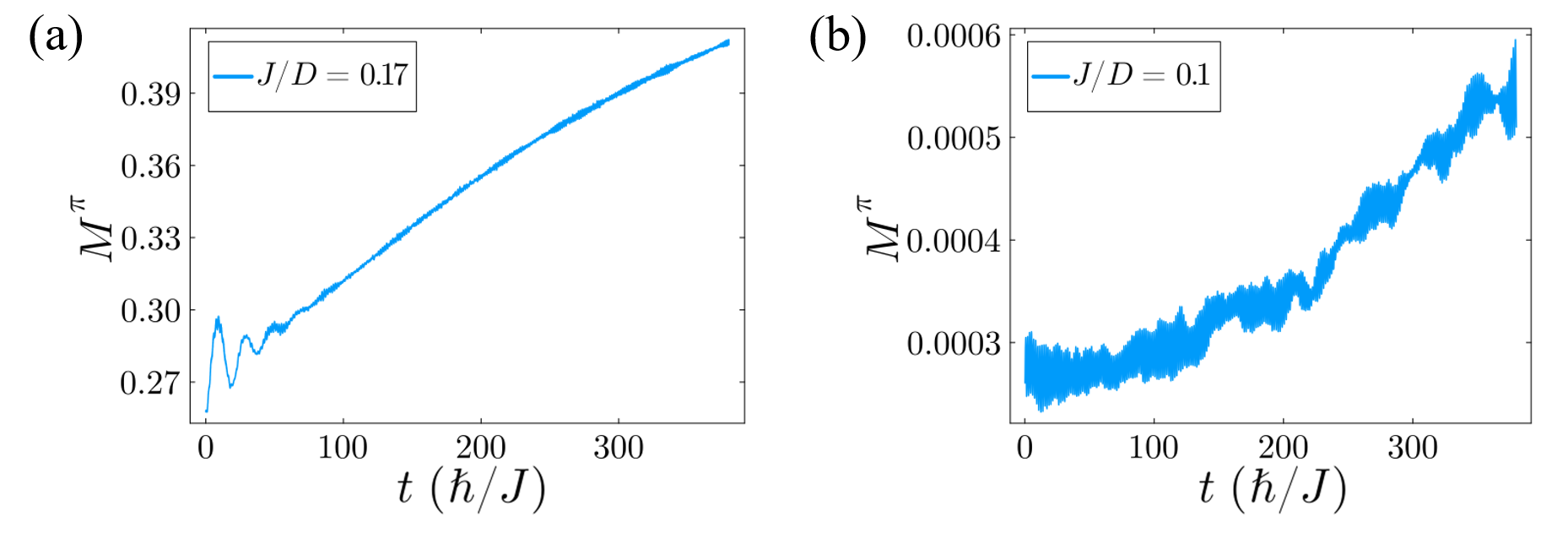}
\caption{Time evolution of the antiferromagnetic magnetization, $M^\pi$, as a function of time averaged over 200 initial states in (a) the AFM phase ($J/D=0.17$, $T/J=0.2$), and (b) the QPM phase ($J/D=0.1$, $T/J=0.2$) the  with drive amplitude $A=0.2$. The data shown here corresponds to the same simulations whose FM $S_z$ magnetization is shown in Fig. \ref{fig:twoPhononDrive}.}
\label{fig:AFMmag--2phonon}
\end{figure}

\section{Effective Hamiltonian in the one-phonon drive}
\label{app:Floquet-one-phonon-drive}

In this section, we compute the Magnus expansion for the single phonon drive in the antiferromagnetic phase for the spin-1 model, with the goal of understanding both the induced AFM order and the difference, at low energy, between the dynamical spin structure factor in the undriven and one-phonon-driven cases. 
Explicitly, as in the main text, we may begin with the Hamiltonian:
\begin{equation}
\label{eq:spin1Ham--1phononDrive}
    H=J\sum_{\langle \br\br'\rangle} \bS(\br)\cdot \bS(\br') + D\sum_\br S^2_z(\br) + H^{\rm sp}_{\rm drive},
\end{equation}
where $H^{\rm sp}_{\rm drive} = -A_{\rm eff} \sum_\br \cos(\Omega t) T_{xz}(\br)$ and $A_{\rm{eff}}$ encompasses the phonon amplitude and spin-phonon coupling strength and $\Omega$ is the phonon frequency, where we have approximated the $xz$ phonon by its sinusoidal drive.
From here, one may easily identify the Fourier components of the time-dependent Hamiltonian:
$$
H^{\rm sp}_{\rm drive,0}=J\sum_{\langle \br\br'\rangle} \bS(\br)\cdot \bS(\br') + D\sum_\br S^2_z(\br),
$$
and
$$
H^{\rm sp}_{\rm drive,+}=H^{\rm sp}_{\rm drive,-}=\frac{A_{\rm eff}}{2}\sum_{\br}T_{xz}(\br).
$$
With a one-phonon drive, we note the the first order expansion outlined in appendix \ref{app:Floquet-two-phonon-drive} is insufficient to realize any effect of the drive, as $H^{\rm sp}_{\rm drive,+} = H^{\rm sp}_{\rm drive,-}$, yielding $H_{\rm eff}^{(1)} = 0$. Thus, we compute the expansion to second order,
\begin{multline}
    H^{(2)}_{\rm{eff}}=\frac{1}{\Omega^2}\left(\sum_{k=1}^\infty \frac{1}{2k} \left( \left[ \left[ H_k, H_0\right], H_{-k} \right] +\left[ \left[ H_{-k}, H_0\right], H_{k} \right] \right)\right) \\
    + \frac{1}{\Omega^2} \Bigg( \sum_{k,m=1}^\infty \frac{1}{3mk} \left( \left[  H_k, \left[H_{m}, H_{-k-m}\right] \right] - \left[  H_k,\left[ H_{-m}, H_{k-m}\right] \right]\right)
    +h.c \Bigg).
\end{multline}
% Explicitly, we may begin with the Hamiltonian:
% \begin{equation}
% \label{eq:spin1Ham--1phononDrive}
%     H=J\sum_{\langle \br\br'\rangle} \bS(\br)\cdot \bS(\br') + D\sum_\br S^2_z(\br) + A_{\rm eff} \sum_\br \cos(\Omega t) T_{xz}(\br),
% \end{equation}
% where $A_{\rm{eff}}$ encompasses the phonon amplitude and spin-phonon coupling strength and $\Omega$ is the phonon frequency, where we have approximated the $xz$ phonon by its sinusoidal drive. 
% From here, one may easily identify the Fourier components of the time-dependent Hamiltonian:
% $$
% H_0=J\sum_{\langle \br\br'\rangle} \bS(\br)\cdot \bS(\br') + D\sum_\br S^2_z(\br),
% $$
% and
% $$
% H_1=H_{-1}=\frac{A_{\rm eff}}{2}\sum_{\br}T_{xz}(\br).
% $$
We thus see that 
% $$
% H^{(0)}_{\rm{eff}}=J\sum_{\langle\br\br'\rangle}S(\br)S(\br')+D\sum_{\br}S_z(\br)^2
% $$
\begin{multline}
H^{(2)}_{\rm{eff}}=-J\frac{A_{\rm eff}^2}{2\Omega^2}\sum_{\langle\br\br'\rangle}\left(S_x(\br)S_x(\br')+4 S_y(\br)S_y(\br')+S_z(\br)S_z(\br')\right) \\
+J\frac{A_{\rm eff}^2}{2\Omega^2}\sum_{\langle\br\br'\rangle}\left(T_{xy}(\br) T_{xy}(\br')+T_{yz}(\br) T_{yz}(\br')+T_{e}(\br) T_{e}(\br')\right)
-D\frac{A_{\rm eff}^2}{4\Omega^2}\sum_\br T_{{e}}(\br) 
\end{multline}
% Ultimately we find:
% $$
% [[H_1,H_0],H_{1}]=-\left(\frac{A^2\alpha}{4}S_x+A^2\beta S_y+\frac{A^2\gamma}{4}S_z+\frac{A^2\Delta}{4} Q_{\rm{eg}} \right)
% $$
where $T_{{e}}(\br)\equiv\sqrt{3}T_{z^2}(\br)-T_{x^2-y^2}(\br)$, and $H^{(0)}_{\rm{eff}}$ is the equilibrium Hamiltonian.
Therefore, we find that the full effective Hamiltonian is given by
\begin{multline}
\label{eq:magnusSpin1AFM-app}
   \! H_{\rm eff} \!\! = \!\! \sum_{\langle \br\br'\rangle} \! J_{\br\br'} (1 \!-\! \frac{A_{\rm eff}^2}{2\Omega^2}) [\bS (\br) \cdot \bS(\br')
   \!+\! (g-1) S_y(\br)S_y(\br')] \\
    \!\! + \! \Gamma \! \sum_{\langle \br\br'\rangle} \! (T_{xy}(\br) T_{xy}(\br')\!+\!T_{yz}(\br) T_{yz}(\br') \!+\! 
   T_{e}(\br) T_{e}(\br')) 
    +\! D (1 \!-\! 3 \frac{A_{\rm eff}^2}{4 \Omega^2})\! \sum_{\br}  S_z^2(\br) \!+\!
   \Delta \! \sum_\br T_{x^2-y^2}(\br)
\end{multline}
where $T_e \!\equiv\! \sqrt{3} T_{z^2}\!-\!T_{x^2\!-\!y^2}$, $T_{z^2} \!=\! 3 S_z^2 \!-\! \bS^2$, $T_{x^2\!-\!y^2} \!=\! S_x^2\!-\!S_y^2$
and the newly generated couplings have strengths
\begin{equation}
    \Gamma \!=\! J \frac{A_{\rm eff}^2}{2\Omega^2};~ g \!=\! \frac{(2 \Omega^2 \!-\! 4 A_{\rm eff}^2)}{(2 \Omega^2\!-\!A_{\rm eff}^2)};~
    \Delta\!=\!D \frac{A_{\rm eff}^2}{4 \Omega^2}.
\end{equation}
% \begin{multline}
% \label{eq:magnusSpin1AFM}
%    H_{\rm eff}\simeq J\sum_{\langle \br\br'\rangle}\left[  \left(1-\frac{5A_{\rm eff}^2}{12\Omega^2}\right) S_x(\br)S_x(\br')+\left(1-\frac{5A_{\rm eff}^2}{3\Omega^2}\right)S_y(\br)S_y(\br')
%    +\left(1-\frac{5A_{\rm eff}^2}{12\Omega^2}\right) S_z(\br)S_z(\br')\right]\\ +D\sum_{\br}\left[ \left(1- \frac{5A_{\rm eff}^2}{4\Omega^2}\right) S_z^2(\br)
%    + \frac{5A_{\rm eff}^2}{12\Omega^2} T_{x^2-y^2}(\br) +\frac{5A_{\rm eff}^2}{6\Omega^2}I\right]  +...
% \end{multline}
One can see here that the single phonon drive acts to renormalize both the exchange constant $J$ and the single-ion anisotropy constant $D$. We may construct a phase diagram for the effective Hamiltonian, $H_{\rm{eff}}=H^{(0)}_{\rm{eff}}+H^{(2)}_{\rm{eff}}$, in order to visualize this effect using a mean field approach, the results are shown in Fig.\ref{fig:MFphaseDiagrams}. Note both the $O(2)$ symmetry breaking when driving the system in the AFM phase and the QPM to $S_y$ AFM transition when driving  the system in the QPM phase. These effects present themselves in both the real time driven spin dynamics shown in Fig. \ref{fig:AFMmag1} and the dynamical spin structure factor (DSSF) data shown in Fig. \ref{fig:spin1flavorwave}(b), wherein the three bands computed for the static case are shifted down as a consequence of the one-phonon drive. 
% Explicitly, using numbers for the parameters $A_{\rm eff}\simeq 5$ and $\Omega \simeq 40$, one can estimate the renormalization of $S_x$ and $S_z$ components of $J$ to be $\sim 0.7\%$, with the renormalization of $D$ and the $S_y$ component of $J$ being $\sim2.1\%$ and $\sim2.8\%$ respectively.

\begin{figure}[h]
\centering
\includegraphics[width=0.95\textwidth]{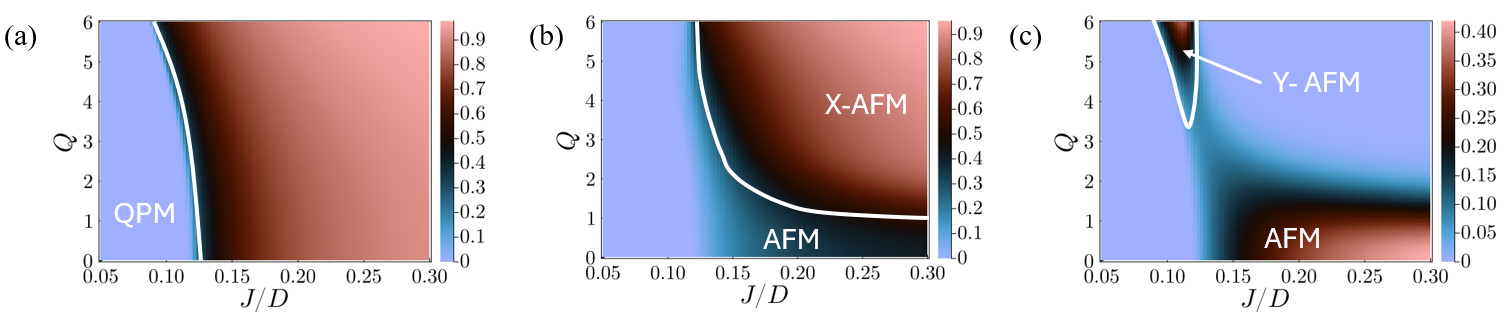}
\caption{Mean field phase diagram of the effective Hamiltonian, $H_{\rm{eff}}=H^{(0)}_{\rm{eff}}+H^{(2)}_{\rm{eff}}$, for the spin 1 model in a single-phonon drive as a function of both $J/D$, the ratio of spin exchange to single-ion anisotropy in the static model, to the steady-state phonon amplitude $Q_0$ which scales effective drive strength $A_{\rm{eff}}$. We highlight both the dipole length, (a), and magnitude of the $S_x$ dipole component, (b), and the magnitude of the $S_y$ dipole component (c). With $Q_{0}=0$ we recover the static case, observing a quantum paramagnetic regime at small $J/D$ that transitions to an easy-plane $xy$ antiferromagnet with increasing $J/D$. Significantly, we note a twofold effect of taking $Q_{0}\neq 0$, firstly that, in the quantum paramagnetic regime, increasing $Q_{0}$ induces a transition into the antiferromagnetic phase, seen in  (a). Secondly, in (b) and (c) we observe the $O(2)$ symmetry breaking seen in the full lattice model (see main text) where driving the system in the AFM regime prefers $S_x$ AFM ordering, while driving the system in the QPM regime favours $S_y$ AFM ordering.}
\label{fig:MFphaseDiagrams}
\end{figure}

\section{Dynamical spin structure factor in the one-phonon drive}
\label{app:DSSFcalc}

\subsection{Calculation of the dynamical spin structure factor}

To study the effect of driving on flavor-wave excitations, we calculate the dynamical spin structure factor (DSSF) in the antiferromagnetic regime. First, we compute the spatial Fourier transform of the spin components
\begin{equation}
    S_\alpha (\mathbf{q}, t) = \frac{1}{\sqrt{N}}\sum_{\mathbf{r}} e^{i\mathbf{q} \cdot \mathbf{r}} S_\alpha (\mathbf{r},t)
\end{equation}
where $S_\alpha (\mathbf{r},t)$ is the classical spin component and $\alpha = x,y,z$. We then calculate the Fourier transform in time as
\begin{equation}
    S_\alpha (\mathbf{q},\omega) = \frac{1}{\sqrt{N_t}}\sum_t e^{i \omega t} S_\alpha (\mathbf{q}, t)
\end{equation}
where $N_t$ is the number of time steps. Finally, since the Fourier transformed spin components are real, then $S_\alpha (\mathbf{q},\omega) = S^{*}_\alpha (-\mathbf{q},-\omega)$ and we can thus calculate the DSSF as
\begin{equation}
    \mathcal{S} (\mathbf{q}, \omega) = \sum_{\alpha} | S_\alpha (\mathbf{q},\omega) |^2
\end{equation}
All DSSF data presented both here and in the main text are multiplied by the quantum to classical correspondence $\frac{\beta \omega}{1-e^{-\beta \omega}}$ \cite{do2023understanding}. 
% We start by calculating the DSSF for our spin-1 model with phonons coupled, but not driven with $\Delta t=0.05/J$, $N_t=7600$ and averaging over 200 initial configurations. We find similar features to those reported in Ref. \cite{do2023understanding}, namely two transverse and one longitudinal mode corresponding to in-plane, out-of-plane and dipolar spin length oscillations, respectively. The gapless Goldstone mode arises from the easy-plane antiferromagnetic interaction, in which global rotations of the lattice in the $xy$ plane are energy conserving. The gapped transverse mode results from the single-ion anisotropy which favors $\langle S_z(\br)\rangle=0$. 
% \begin{figure}[h]
% \centering
% \includegraphics[width=0.45\textwidth]{Figures/SM-figS7.png}
% \caption{Flavor-wave dispersion, calculated as $\mathcal{S}(\mathbf{q},\omega)$, in the AFM phase ($J/D=0.4$) averaged over 200 configurations to a maximum time $t=200/J$, with sample spacing  $dt=0.1/J$. There are two transverse modes, one which is Goldtsone-like and one which is gapped, and one longitudinal mode. See Fig. 4 in the main text for more details.}
% \label{fig:DSSF-withPhonons}
% \end{figure}

\subsection{The effect of spin-phonon coupling and resonant phonon drive}

In this section, we present additional plots of the structure factor, highlighting the role that spin-phonon coupling plays in the spin-wave modes. Absent any spin-phonon coupling ($\lambda/J = 0$) with $\Delta t=0.05/J$, $N_t=3000$ and averaging over 200 initial configurations, the DSSF gives very similar results to those presented in Ref \cite{do2023understanding}. 
% We start by calculating the DSSF for our spin-1 model with phonons coupled, but not driven with $\Delta t=0.05/J$, $N_t=7600$ and averaging over 200 initial configurations. We find similar features to those reported in Ref. \cite{do2023understanding}, 
% Namely two transverse and one longitudinal mode corresponding to in-plane, out-of-plane and dipolar spin length oscillations, respectively. 
% The gapless Goldstone mode arises from the easy-plane antiferromagnetic interaction, in which global rotations of the lattice in the $xy$ plane are energy conserving. The gapped transverse mode results from the single-ion anisotropy which favors $\langle S_z(\br)\rangle=0$.  
In particular, in Fig. \ref{fig:DSSF--SM}(a), two transverse and one longitudinal mode corresponding to in-plane, out-of-plane and dipolar spin length oscillations, respectively. The gapless Goldstone mode arises from the easy-plane antiferromagnetic interaction, in which global rotations of the lattice in the $xy$ plane are energy conserving. The gapped transverse mode results from the single-ion anisotropy which favors $\langle S_z(\br)\rangle=0$. 
% The Goldstone mode shows no gap, and exhibits an energy of $\omega \approx 9.0 J \approx 2.4 $meV at $q=(0,0,0)$ and $q=(2\pi,2\pi,2\pi)$. We also see the longitudinal mode and the gapped transverse mode at similar energy scales. 
With the addition of spin-phonon coupling in equilibrium ($A=0, \lambda/J=3.5$), the spin-wave modes are significantly renormalized such that they all are shifted to lower energies, however the Goldstone mode remains gapless (Fig. \ref{fig:DSSF--SM}(b)). Further, with the inclusion of the drive of a single phonon mode, the spin waves are again further shifted to lower energies, and a gap arises in the Goldstone mode at $q=(\pi,\pi,\pi)$ (Fig. \ref{fig:DSSF--SM}(c)), which we have explained using the Floquet-Magnus expansion described previously. We also note a further splitting of the Goldstone and longitudinal bands at $q=(0,0,0)$ and $q=(2\pi,2\pi,2\pi)$.

\begin{figure}[h]
\centering
\includegraphics[width=1.0\textwidth]{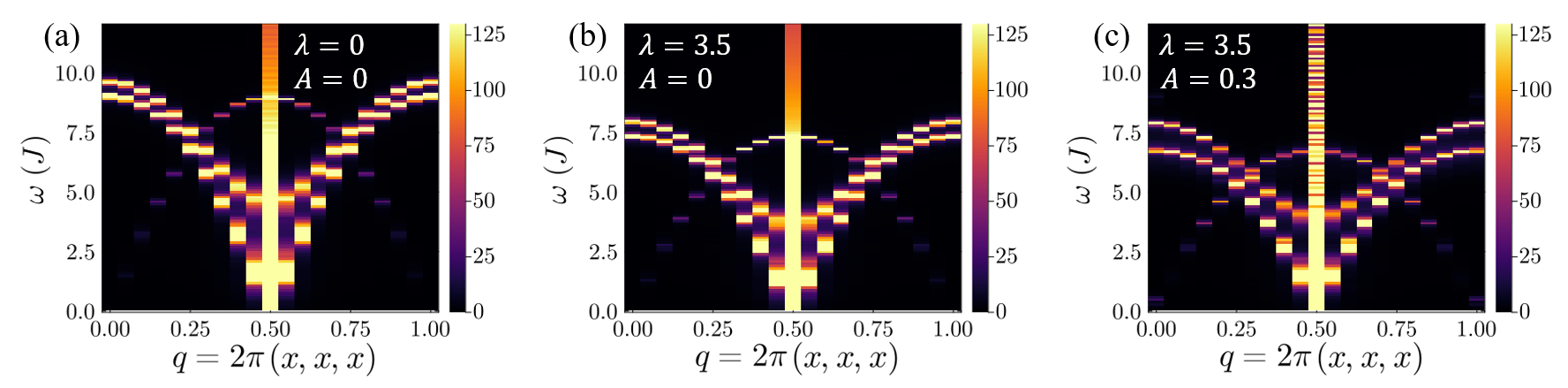}
\caption{Flavor-wave dispersion, calculated as $\mathcal{S}(\mathbf{q},\omega)$, in the AFM phase ($J/D=0.17$) averaged over 200 configurations to a maximum time $t=150/J$, with sample spacing  $\Delta t=0.05/J$. The DSSF is shown for three different scenarios: (a) the spin system with no spin-phonon coupling, (b) the coupled spin and phonon system in equilibrium (ie. with no phonon drive), and (c) the coupled spin and phonon system with a drive of strength $A=0.3$. In each scenario, we can identify two transverse modes, one which is Goldstone-like and one which is gapped, and one longitudinal mode. We see that coupling the phonons to the system causes a downward shift of the spin-wave modes, and the phonon drive causes a further shift.}
\label{fig:DSSF--SM}
\end{figure}

\subsection{Phonon-induced Goldstone gap}

We identify the gap at $\omega \sim 0.4J$, determined from a feature in the DSSF at $q=(\pi,\pi,\pi)$, as described in the main text. We further confirm the existence of the gap using a fit to the Goldstone mode in both the undriven and driven systems. Using the peaks in the DSSF which correspond to the Goldstone mode, we fit the data to the function
% The points indicate the peaks in the DSSF corresponding to the Goldstone mode, while the lines correspond to a fit to the function
$$
\sqrt{p_0^2 + (p_1\cos(q\pi) + p_2\cos(3q\pi))^2}
$$
where $p_0$, $p_1$, and $p_2$ are fitting parameters. The fits are shown in Fig. \ref{fig:DSSFfit}, where the points indicate the peaks in the DSSF, while the lines correspond to the fit. As is clear, there arises a gap of $\omega = 0.34J$ in the driven system, further confirming that the $O(2)$ symmetry breaking induced by the one-phonon drive gaps out the Goldstone mode.
% which gives a gap of $\omega = 0.34J$ in the driven system.

\begin{figure}[h]
\centering
\includegraphics[width=0.4\textwidth]{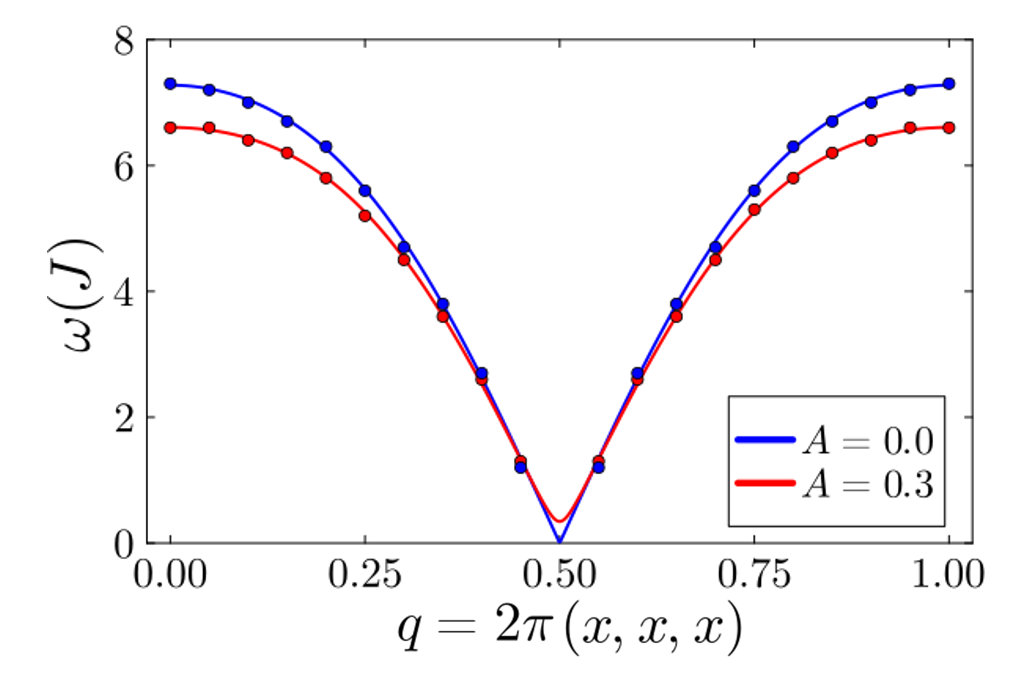}
\caption{Fit to the dynamical spin structure factor of the Goldstone mode for the system under a constant periodic drive (red) and the system in equilibrium (blue). The points indicate the peaks in intensity of the DSSF corresponding to the Goldstone mode, while the lines indicate the fitting function to those points. It is clear that with a drive strength of $A=0.3$, the Goldstone mode is gapped at $q = (\pi,\pi,\pi)$.}
\label{fig:DSSFfit}
\end{figure}

\subsection{High energy Floquet sidebands}

Finally, Fig.~\ref{fig:spin1flavorwave-HighE} plots $\mathcal{S}_(\bq,\omega)$ for a one-phonon drive in the AFM phase, for higher energies
in the vicinity of the phonon energy $\Omega/J \!\sim\! 40$, showing clear signatures of phonon-induced Floquet copies 
of the flavor-wave spectrum as well as a signature of the 
flat phonon mode itself in the magnetic $\mathcal{S}(\bq,\omega)$. 
The higher energy Floquet modes have lower intensity, and
show up as both $\pm$ branches of the dispersion centered around the phonon energy, since inversion symmetry
fixes $\mathcal{S}(\bq,-\omega)=\mathcal{S}(\bq,\omega)$, and ignoring spectral intensities 
$\mathcal{S}(\bq,-\omega) \sim \mathcal{S}(\bq,\Omega-\omega)$.
One aspect of the higher energy Floquet modes we have not understood 
is the difference in dispersion of the Goldstone mode in the higher energy Floquet spectrum when compared with the low energy 
regime. We tentatively attribute this difference to phonon dissipation effects but 
hope to clarify this in future work. These Floquet flavor-wave 
excitations, which are driven dipolar-quadrupolar modes, could potentially be detected for $\bq=0$
using THz spectroscopy. A more challenging experiment, analogous to time and angle resolved photoemission
spectroscopy used to observe electronic Floquet states in topological insulators and graphene \cite{floquet_trARPES_gedik_2013,floquet_trARPES_mathias_2024,floquet_trARPES_gedik_2024},
would be inelastic neutron scattering in the presence of driven phonons \cite{Pump_neutron_2024}.

\begin{figure}[h]
\centering
\includegraphics[width=0.5\textwidth]{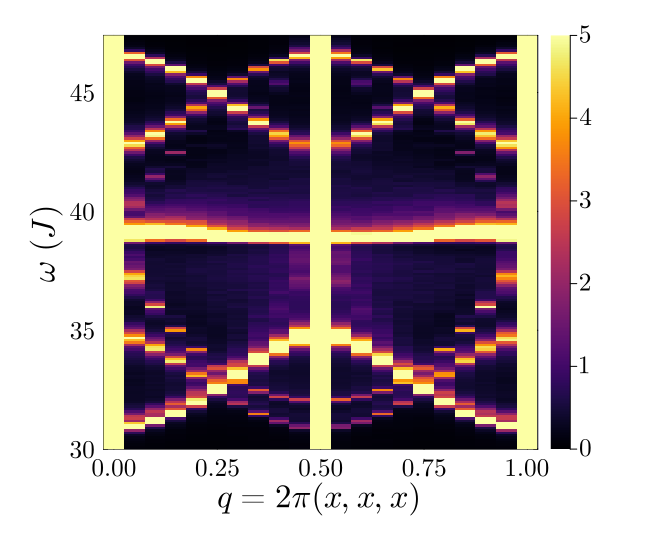}
\caption{Flavor-wave dispersion, calculated as $\mathcal{S}(\mathbf{q},\omega)$, in the AFM phase ($J/D=0.17$) at energies in the vicinity of the phonon frequency $\Omega/J \approx 40$, averaged over 200 configurations to a maximum time $t=150/J$.}
\label{fig:spin1flavorwave-HighE}
\end{figure}

\end{document}